\documentclass{article}

\usepackage{amsmath}
\usepackage{psfrag}
\usepackage{geometry}
\usepackage[normalem]{ulem}
\usepackage[dvips]{graphicx,color}

\begin{document}

\title{Local Momenta and a Three-Body Gauge}
\author{Michael Jay Schillaci,\emph{Francis Marion University, Florence, SC 29501}}
\date{\today}
\maketitle

\begin{abstract}

In recent years researchers have attempted to improve the
continuum state three-body wavefunction for three, mutually
interacting Coulomb particles by including, so called,
\emph{local momentum} effects, which depend upon the logarithmic
gradient of the continuum, two-body Coulomb waves. Using the
\emph{exact} three-body wavefunction in the region where two of
the three particles remain close, a revised description of these
local momenta, is attained and predicts that a quantum-mechanical
impulse may develop in the reaction zone, causing
like-sign--charged particles to decrease their radial separation
and  opposite-sign--charged particles to increase their radial
separation. The consequences of these predictions are investigated
through both quantum and semi-classical techniques where the
total energy of a two-body continuum Coulomb system in the
presence of a third, mutually interacting body are analyzed.
Numerical calculations confirm that while ignoring these local
effects for \emph{light-ion--atom} processes, may be appropriate,
three-body effects may dominate in the reaction zone for
\emph{heavy-ion--atom} processes.  The techniques developed here
are then applied to explain the observed asymmetry in the data
collected by Wiese, \emph{et. al.},[\textbf{PRL} 79, 4982] on the
correlated breakup of three massive, Coulomb interacting
particles. The results are attributed to a genuine three-body
effect which rearranges particles in the reaction zone, while
retaining the appropriate asymptotic behavior. Preliminary
calculations show a deviation of less than 10\% between the
predicted and observed asymmetries. The strength of these results
is then used to argue that the local momenta, herein developed,
be treated as a formal gauge constraint for three-body
interactions. This hypothesis is investigated and it is shown
that a real-valued, position-dependent phase is added to the
wavefunction. A semi-classical analysis of the proposed
three-body gauge, reveals that while genuine three-body effects
may arise in the reaction zone, the asymptotic form of the
relevant two-body Hamiltonian remains unchanged for relative
energies greater than $\sim 1\mathrm{eV}$ for all atomic species.
Further analysis shows however that one may detect asymptotic
variations in the scattering amplitudes for massive systems at
energies $\sim \mathrm{\mu eV}$. These results provide
convincing theoretical and physical evidence for the success of
many current experiments and indicate that more experimentation
with near-threshold, massive three-body systems is needed.
\end{abstract}


\section{Introduction}
The "three-body" problem is as old as the study of Physics itself.
After Sir Isaac Newton showed (in his \emph{Principia}) that it was
possible to infer the orbits of \emph{two}, mutually interacting
bodies using only the laws of mathematics, mankind has endeavored
to derive an analytic description of the motion of \emph{three},
mutually interacting particles.

Three-body interactions abound in natural processes as diverse as stellar evolution
and thin-film growth and occur over a very wide range of energies. For the present
discussion, continuum atomic scattering will be considered for three mutually
interacting charged particles interacting via the infinite range Coulomb potential.
Calculation of local distortion effects herein derived will be carried out over
representative energies of between $3-54.4\ \mathrm{eV}$.\footnote{These energies
were chosen because of the particular relevance to the results offered in \cite{BBK}
on electron or positron scattering from hydrogen.}

In Section \ref{sec:notation}, the traditional Jacobi coordinates
are introduced and key historical results are given. While the
notation used  here is not substantially different, minor errors
have been corrected and additional properties and unique results have been
added.

A revised description of the so called local momenta, first
presented in \cite{AM} is offered in Section \ref{sec:origin}.
While the local momenta derived here also depend upon the
logarithmic gradient of a continuum state, Coulomb wave - here
referred to as the local distortion - the coordinate dependence
is such that variations of the momenta can not be ignored in an
\emph{a priori} manner. The development continues in Section
\ref{sec:general}, where a detailed discussion of the mathematical
and physical properties of the local momenta is offered. Appendix
\ref{app:1f1} shows that the local distortion can be expressed as
a damped oscillator function that is analytic over all regions,
thus improving the ability to assess the possible contributions of
these local effects in both the inner (reaction-zone), and the
outer (asymptotic) regions of the three-body scattering event.
\footnote{While there is no rigorous convention for the
quantification of these regions, the ``reaction zone'' is here
taken to be the region wherein $r < 10 a_0$.} This analysis
reveals that while local distortion effects generally
``fall-off'' in the asymptotic regions, they may alter the
outcome of a scattering event by rearranging particles in the
reaction-zone. Appendix \ref{app:kc5c} demonstrates however that
one may incorporate local effects while retaining the appropriate
asymptotic form for the continuum state, three-body wavefunction.
From this a new interpretation of the three-body scattering event
emerges, wherein a two-body continuum pair acquires a local
momentum by scattering off of an exact interaction potential.

Section \ref{sec:rearrange} illustrates that generalized local
momentum effects may be used to explain the observed asymmetry in
the data collected in the triple-coincidence detection experiment
of Wiese et. al., \cite{Wiese}. The experiment considered three
large, very nearly equal mass, mutually interacting, charged
particles. With the appropriate total center-of-mass energy and
reduced mass parameters chosen to reflect those used in the
actual experiment, the predicted local effects are shown to be
large enough to reproduce the observed asymmetry. This result is
taken as motivation to hypothesize that the generalized local
momentum be treated as a possible gauge transformation for
three-body interactions. In Sec. \ref{sec:gauge} this hypothesis
is formalized and a semi-classical analysis reveals that while a
real-valued, position dependent phase is added to the two-body
continuum state wavefunction, the asymptotic form of the relevant
two-body Hamiltonian remains unchanged.

\section{Notation}
\label{sec:notation}

\begin{figure}
\begin{center}
\fbox{$\text{The Jacobi Coordinates}$}\\ \psfrag{ra}{\Large
$\vec{r}_\alpha$} \psfrag{rb}{\Large $\vec{r}_\beta$}
\psfrag{rc}{\Large $\vec{r}_\gamma$} \psfrag{pa}{\Large
$\vec{\rho}_\alpha$} \psfrag{pb}{\Large $\vec{\rho}_\beta$}
\psfrag{pc}{\Large $\vec{\rho}_\gamma$} \psfrag{ma}{\Large
$m_\alpha$} \psfrag{mb}{\Large$m_\beta$}
\psfrag{mc}{\Large$m_\gamma$}
\resizebox{3in}{!}{\includegraphics{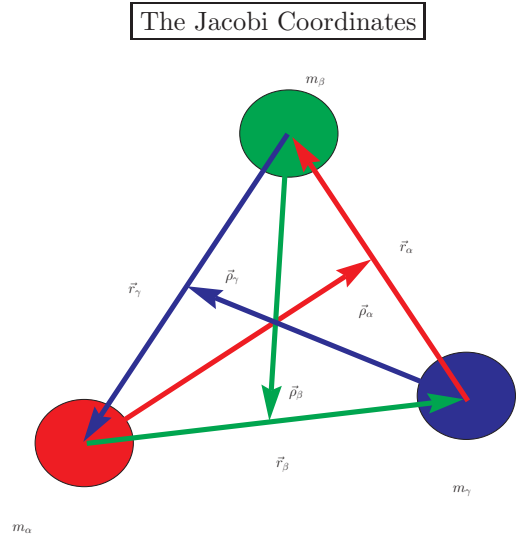}}
\end{center}
\caption{The Jacobi coordinates are most often used to study many-body kinematics,
because any orthogonal pair, $(\vec{r}_\nu,\vec{\rho}_\nu),
\quad \nu=\alpha,\beta,\gamma$ may be used. See equations (\ref{eq:masses}) for the
definition of the reduced masses.}
\label{fig:jacobi}
\end{figure}

The traditional Jacobi coordinates, here denoted by $(\vec{r}_\nu,
\vec{\rho}_\nu)$ along with their respective, congugate momenta, $(\vec{k}_\nu,
\vec{q}_\nu),$ where $\nu=\alpha,\beta,\gamma$ will be used to indicate a particular channel
representation. These coordinates are particularly well suited for the study of the
motions of three mutually interacting particles because they locate the
conventional\footnote{While there is some debate in the literature, as noted in
\cite{Fiziev}, over the appropriate reduced masses to use in the Classical treatment
of the three-body problem, the possible renormalization of mass in the Quantum
treatment makes these arguments irrelevant here.} reduced masses of the system. These
coordinates are shown in Figure \ref{fig:jacobi} and the ``alpha-channel''
representation will be used unless otherwise specified.

The reduced masses located by the $(\vec{r}_\alpha,\vec{\rho}_\alpha)$ coordinates
are given by
\begin{subequations}\label{eq:masses}
\begin{eqnarray}
\mu_\alpha&\equiv&\frac{m_\beta m_\gamma}{m_\beta + m_\gamma}\\
M_\alpha&\equiv&\frac{m_\alpha\left(m_\beta + m_\gamma\right)}{m_\alpha + m_\beta +
m_\gamma},
\end{eqnarray}
\end{subequations}
respectively and the other masses are defined cyclically.

In addition, there are two relationships that the Jacobi coordinates and their
conjugate momenta obey that will be of use in the current development. These are,
\begin{subequations}
\begin{eqnarray}
&\vec{r}_\alpha+\vec{r}_\beta+\vec{r}_\gamma=0&\label{eq:relationa}\\
&\frac{\vec{k}_\alpha}{\mu_\alpha}+\frac{\vec{k}_\beta}{\mu_\beta}+
\frac{\vec{k}_\gamma}{\mu_\gamma}=0.&\label{eq:relationb}
\end{eqnarray}
\end{subequations}
While (\ref{eq:relationa}) can be ``seen'' in Figure \ref{fig:jacobi},
(\ref{eq:relationb}) is more subtle and is a statement of the ``relative-velocity
conservation'' for the three-body system. These relationships follow from the
orthogonality of the Jacobi coordinates and are easily verified using the
transformation matrices for the coordinates,
\begin{subequations}
\begin{eqnarray}
&\left(\begin{array}{c} \vec{\rho}_\beta\\ \vec{r}_\beta
\end{array}\right)=
\left(\begin{array}{cc} -\frac{\mu_\beta}{m_\gamma}&-\frac{\mu_\alpha}{M_\beta}\\
                1&-\frac{\mu_\alpha}{m_\gamma}
\end{array}\right)
\left(\begin {array}{c} \vec{\rho}_\alpha\\ \vec{r}_\alpha
\end{array}\right),&\label{eq:coordinatea}\\
&\left(\begin {array}{c} \vec{\rho}_\gamma\\ \vec{r}_\gamma
\end{array}\right)=
\left(\begin{array}{cc} -\frac{\mu_\gamma}{m_\beta}&\frac{\mu_\alpha}{M_\gamma}\\
                -1&-\frac{\mu_\alpha}{m_\beta}
\end{array}\right)
\left(\begin {array}{c} \vec{\rho}_\alpha\\ \vec{r}_\alpha
\end{array}\right),&\label{eq:coordinateb}
\end{eqnarray}
\end{subequations}
and for the momenta,
\begin{subequations}
\begin{eqnarray}
&\left(\begin {array}{c} \vec{q}_\beta\\ \vec{k}_\beta
\end{array}\right)=
\left(\begin{array}{cc} -\frac{\mu_\alpha}{m_\gamma}&-1\\ \frac{\mu_\beta}{M_\alpha}
&-\frac{\mu_\beta}{m_\gamma}
\end{array}\right)
\left(\begin {array}{c} \vec{q}_\alpha\\ \vec{k}_\alpha
\end{array}\right),&\label{eq:momentuma}\\
&\left(\begin {array}{c} \vec{q}_\gamma\\ \vec{k}_\gamma
\end{array}\right)=
\left(\begin{array}{cc} -\frac{\mu_\alpha}{m_\beta}&1\\
-\frac{\mu_\gamma}{M_\alpha}&-\frac{\mu_\gamma}{m_\beta}
\end{array}\right)
\left(\begin {array}{c} \vec{q}_\alpha\\ \vec{k}_\alpha
\end{array}\right).&\label{eq:momentumb}
\end{eqnarray}
\end{subequations}

A further consequence of these transformations is that the three-body, Coulomb
potential may be expressed as follows\footnote{While it is conventional to use
``atomic units'' wherein, $\hbar=e=1$, all physical constants are retained so that
the interested reader may verify explicit numerical results cited later in the text
without normalization.}:
\begin{equation}
V_C^\alpha(\vec{r}_\alpha,\vec{\rho}_\alpha)=\frac{e^2}{4\pi\epsilon_0}
\left(\frac{Z_\beta Z_\gamma}{r_\alpha}+
\underbrace{\frac{Z_\alpha Z_\gamma}
{\left|\vec{r}_\alpha-\frac{\mu_\alpha}{m_\gamma} \vec{\rho}_\alpha \right|}}_{r_\beta}+
\underbrace{\frac{Z_\alpha Z_\beta}
{\left|\vec{r}_\alpha+\frac{\mu_\alpha}{m_\beta} \vec{\rho}_\alpha
\right|}}_{r_\gamma}
\right),
\end{equation}
where the $Z_\nu$ ($\nu=\alpha,\beta,\gamma$) are the appropriate
charge signs. Moreover, the orthogonality of the Jacobi
coordinates can be used to show that, the three-body
Schr\"odinger equation may be written in a channel-independent
way. i.e.,
\begin{equation}
\left(H_0^\nu+V_C^\nu\right)\Psi_{\vec{k}_\nu,\vec{q}_\nu}(\vec{r}_\nu,\vec{\rho}_\nu)=
{\mathcal{E}}_t \Psi_{\vec{k}_\nu,\vec{q}_\nu}(\vec{r}_\nu,\vec{\rho}_\nu),
\end{equation}
where
\begin{equation}
H_0^\nu=-\frac{\hbar^2}{2\mu_\nu}\vec{\nabla}^2_{\vec{r}_\nu}-
\frac{\hbar^2}{2M_\nu}
\vec{\nabla}^2_{\vec{\rho}_\nu},
\end{equation}
and $V_C^\nu(\vec{r}_\nu,\vec{\rho}_\nu)$ has been used, with supressed coordinate
dependence, to indicate the full three-body Coulomb potential of the $\nu$-channel.
Here ${\mathcal{E}}_t$ is the total center of mass energy of three-body system.

To date, the most successful and widely used approximate solution
to the three-body Schr\"odinger equation is the paradigm ``3C''
wavefunction proposed by Redmond\cite{Redmond},\cite{Rosenberg}
and rigorously derived and tested by Brauner \emph{et.
al.}\cite{BBK}. The solution is valid in the asymptotic region
where all particles are far apart. Traditionally one denotes this
region with, $\Omega_0$, and the solution is given by,
\begin{equation}\label{eq:redmond}
\Psi_{\vec{k}_\nu,\vec{q}_\nu}^{RED(\Omega_0)}
(\vec{r}_\nu,\vec{\rho}_\nu)=
e^{i(\vec{k}_\nu\cdot\vec{r}_\nu+\vec{q}_\nu\cdot\vec{\rho}_\nu)}
\prod_{\nu=\alpha,\beta,\gamma} e^{-i\eta_{k_\nu}\ln\zeta_{k_\nu}},
\end{equation}
where for instance,
\begin{equation}
\eta_{k_\alpha}\equiv
Z_\beta Z_\gamma\left(\frac{e^2}{4 \pi \epsilon_0
\hbar}\right)\frac{\mu_\alpha}{\hbar k_\alpha}
\end{equation}
is the atomic Sommerfeld parameter. The hyper-parabolic coordinate,
\begin{equation}
\zeta_{k_\nu}\equiv k_\nu r_\nu+\vec{k}_\nu\cdot\vec{r}_\nu,
\end{equation}
has been introduced for notational simplicity only, and with the above definition,
the wavefunction (\ref{eq:redmond}) satisfies all incoming boundary conditions in the
region $\Omega_0$. The overwhelming opinion in the literature is that all valid
three-body wavefunctions must match smoothly with this solution in the region
$\Omega_0$.

The logarithmic phase factors in the Redmond solution are present \emph{physically}
because of the infinite nature of the Coulomb potential, and arise
\emph{mathematically} as the leading term in the asymptotic expansion of the
confluent hypergeometric function. i.e.,\cite{GR}
\begin{subequations}
\begin{eqnarray}
C_{\eta_{k_\nu}}(\zeta_{k_\nu}) &=&N_{\eta_{k_\nu}}
{}_{1}F_{1}(i\eta_{k_\nu},1,-i\zeta_{k_\nu})\\
C_{\eta_{k_\nu}}(\zeta_{k_\nu})&=&\Gamma[1-i\eta_{k_\nu}]e^{-\frac{\eta_{k_\nu}\pi}{2}}\sum_{n=0}^{\infty}\frac{(i\eta_{k_\nu})_n}{(1)_n}\frac{(-i\zeta_{k_\nu})^n}{n!}\\
C_{\eta_{k_\nu}}(\zeta_{k_\nu})&=&\lim_{{r_\nu}\to\infty}e^{-i\eta_{k_\nu}\ln\zeta_{k_\nu}}.
\end{eqnarray}
\end{subequations}

These ``C-functions'' provide one of the representations for the \emph{exact}
solution to the \emph{two-body} scattering problem. That is, if one writes the
two-body wavefunction as
\begin{equation}
\Psi_{\vec{k}_\nu}^{Two-Body}(\vec{r}_\nu)=e^{i\vec{k}_\nu\cdot\vec{r}_\nu}
C_{\eta_{k_\nu}}(\zeta_{k_\nu}),
\end{equation}
then substitution of this form into the two-body Schr\"odinger equation shows that
\begin{equation}\label{eq:cka}
\left[-\frac{\hbar^2}{2\mu_\alpha}\vec{\nabla}^2_{\vec{r}_\alpha}-
i\frac{\hbar}{\mu_\alpha}\vec{k}_\alpha\cdot\vec{\nabla}_{\vec{r}_\alpha}
+\frac{Z_\beta Z_\gamma e^2}{4 \pi
\epsilon_0}\frac{1}{r_\alpha}\right]
C_{\eta_{k_\alpha}}(\zeta_{k_\alpha})=0.
\end{equation}

While the very intuitive solution (\ref{eq:redmond}) has been used with great success
by researchers to model both atom-ion \cite{BBK},\cite{Jones} and photo-ionization
processes \cite{Qiu}, a more robust solution has been sought in recent
years\cite{AM},\cite{Lieber}. Particularly, a solution is sought that may be extended
into the so called ``interior regions,'' where at least two of the three particles
remain close to one another. It is significant to note that all of the successes of
the 3C wavefunction have been achieved for \emph{light-ion - atom} systems, where the
asymptotic form in $\Omega_0$ has been shown to be generally adequate. What is sought
however is a more precise accounting of the intimacies of the three-body interaction
for arbitrary masses and in all regions of the scattering space.

\section{Origin of the Local Momenta}
\label{sec:origin}

Because an exact solution to the atomic three-body problem does not exist, the best
hope in achieving an improved wavefunction has been to improve the approximation
schemes used. Generally, these approximation schemes fall into two categories:
\begin{itemize}
\item Approximating the Kinetic Terms of the Hamiltonian (i.e., the Eikinol approximation)
\item Approximating the Potential Terms of the Hamiltonian
\end{itemize}
While these techniques would seem to be mutually exclusive, the
problem is that the inseparable nature of the three-body Coulomb
potential makes the choice of the kinematic description
impossible. Research has thus continued along a fragile path and
to account for approximation and/or distortion effects, two well
established interpretations have emerged:
\begin{itemize}
\item Introduce a \emph{Local Momentum}, which depends upon the radial separation of
two of the three particles through the logarithmic gradient of the continuum two-body
wavefunction, and attribute the distortion to a velocity-dependent, auxilary
potential. cf., \cite{AM}
\item Introduce an \emph{Effective Charge}, which depends upon the radial separation of
two of the three particles through the logarithmic gradient of the continuum two-body
wavefunction, and attribute the distortion to dynamical screening of charges. cf.,
\cite{Berakdar}
\end{itemize}

The common dependence upon the two-body solution in these two interpretations is
clear and in both the dependence is derived in a rigorous, but \emph{a posteriori}
way, to satisfy the relevant boundary conditions in the asymptotic regions. The
natural question is if this dependence can be achieved in an \emph{a priori} way,
utilizing only physical and mathematical intuition.

To approach an answer to this question, it is important to know that there does exist
an exact solution to the continuum three-body Schr\"odinger equation for the case
when two of the particles remain close together(or equivalently if one of the
particles is infinitely massive). The solution depends critically upon the form of
the three-body Coulomb potential in the region $\Omega_\alpha$, where two of the
particles, $\beta$ and $\gamma$ are close together, while particle $\alpha$ is
infinitely far away from both of them. That is, one may write a description of the
region $\Omega_\alpha$ succinctly as follows:
$$\Omega_\alpha:\quad \lim{r_\alpha\to\infty}, \quad \lim{\frac{r_\alpha}{\rho_\alpha}\to\ 0}$$
Clearly the region $\Omega_\alpha$ matches smoothly with the region $\Omega_0$, and
to investigate the behavior of the three-body Coulomb potential in this region, one
may write
\begin{equation}
V_C^{\alpha}( \vec{r}_\alpha,\vec{\rho}_\alpha)=
\frac{Z_\beta Z_\gamma e^2}{4\pi\epsilon_0}\frac{1}{r_\alpha}+
\frac{Z_\alpha e^2}{4\pi\epsilon_0}\frac{1}{\rho_\alpha}\times
\sum_{L'=0}^{\infty}Z_{L'}\left(\frac{r_\alpha}{\rho_\alpha}\right)^{L'}
P_{L'}(\hat{r}_\alpha\cdot\hat{\rho}_\alpha),\quad r_\alpha\ll \rho_\alpha,
\end{equation}
where
$$Z_{L'}=Z_\beta\left(-\frac{\mu_\alpha}{m_\beta}\right)^{L'}+
Z_\gamma\left(\frac{\mu_\alpha}{m_\gamma}\right)^{L'},$$ and the
$P_{L'}(\hat{r}_\alpha\cdot\hat{\rho}_\alpha)$ are the Legendre polynomials of the
first kind.

Now in the region $\Omega_\alpha,$ the only surviving term in the expansion is for
$L'=0.$ Hence the potential has the following asymptotic form:
\begin{equation}
V_C^{\alpha\ (\Omega_\alpha)}(\vec{r}_\alpha,\vec{\rho}_\alpha)=
\frac{Z_\beta Z_\gamma e^2}{4\pi\epsilon_0}\frac{1}{r_\alpha}+
\frac{Z_\alpha (Z_\beta + Z_\gamma) e^2}{4\pi\epsilon_0}\frac{1}{\rho_\alpha}.
\end{equation}
The second term in this expansion is often referred to as the ``reduced charge
potential,'' and the resulting form of the three-body wavefunction is separable.
e.g., the three-body Schr\"odinger equation takes the asymptotic form,
\begin{equation}
\left[-\frac{\hbar^2}{2\mu_\alpha}\vec{\nabla}^2_{\vec{r}_\alpha}
+\frac{Z_\beta Z_\gamma e^2}{4 \pi \epsilon_0}\frac{1}{r_\alpha}
-\frac{\hbar^2}{2M_\alpha}\vec{\nabla}^2_{\vec{\rho}_\alpha}
+\frac{Z_\alpha(Z_\beta+Z_\gamma) e^2}{4 \pi \epsilon_0}\frac{1}{\rho_\alpha}\right]
\Psi_{\vec{k}_\alpha,\vec{q}_\alpha}^{\Omega_\alpha}
(\vec{r}_\alpha,\vec{\rho}_\alpha)= {\mathcal{E}}_t
\Psi_{\vec{k}_\alpha,\vec{q}_\alpha}^{\Omega_\alpha}(\vec{r}_\alpha,\vec{\rho}_\alpha).
\end{equation}
Because the term $\frac{Z_\alpha (Z_\beta + Z_\gamma) e^2}{4\pi\epsilon_0
\rho_\alpha}$ couples to the kinetic term
$-\frac{\hbar^2}{2M_\alpha}\vec{\nabla}_{\vec{\rho}_\alpha}^2$, the three-body
Hamiltonian naturally separates and if one assumes that the three-body wavefunction
has the standard plane wave form,
\begin{equation}
\Psi_{\vec{k}_\alpha,\vec{q}_\alpha}^{\Omega_\alpha}(\vec{r}_\alpha,\vec{\rho}_\alpha)=
e^{i(\vec{k}_\alpha\cdot\vec{r}_\alpha+\vec{q}_\nu\cdot\vec{\rho}_\nu)}
\Phi_{\vec{k}_\alpha,\vec{q}_\alpha}(\vec{r}_\alpha,\vec{\rho}_\alpha),
\end{equation}
where the total energy of the three-body system may be written as
\begin{equation}
{\mathcal{E}}_t = \frac{\hbar^2 k_\alpha^2}{2 \mu_\alpha}+
\frac{\hbar^2 q_\alpha^2}{2 M_\alpha}>0,
\end{equation}
for continuum scattering, then substitution into the asymptotic three-body
Schr\"odinger equation yields the following equation for
$\Phi_{\vec{k}_\alpha,\vec{q}_\alpha}(\vec{r}_\alpha,\vec{\rho}_\alpha)$ in the
region $\Omega_\alpha$:
\begin{equation}\label{eq:exactschrodinger}
\left[-\frac{\hbar^2}{2\mu_\alpha}\vec{\nabla}^2_{\vec{r}_\alpha}
+\frac{Z_\beta Z_\gamma e^2}{4 \pi \epsilon_0}\frac{1}{r_\alpha}
-\frac{\hbar^2}{2M_\alpha}\vec{\nabla}^2_{\vec{\rho}_\alpha}
+\frac{Z_\alpha(Z_\beta+Z_\gamma) e^2}{4 \pi \epsilon_0}\frac{1}{\rho_\alpha}\right]
\Phi_{\vec{k}_\alpha,\vec{q}_\alpha}
(\vec{r}_\alpha,\vec{\rho}_\alpha)= 0.
\end{equation}
Therefore the exact solution in the region $\Omega_\alpha$ is given by,
\begin{equation}\label{eq:exact}
\Psi_{\vec{k}_\alpha,\vec{q}_\alpha}^{(\Omega_\alpha)}
(\vec{r}_\alpha,\vec{\rho}_\alpha)=
e^{i(\vec{k}_\alpha\cdot\vec{r}_\alpha+\vec{q}_\nu\cdot\vec{\rho}_\nu)}
C_{\eta_{k_\alpha}}(\zeta_{k_\alpha}) C_{\eta_{q_\alpha}}(\zeta_{q_\alpha}).
\end{equation}
Here the function $C_{\eta_{q_\alpha}}(\zeta_{q_\alpha})$ is a reduced charge
continuum state Coulomb wave, which satisfies
\begin{equation}\label{eq:cqa}
\left[-\frac{\hbar^2}{2M_\alpha}\vec{\nabla}^2_{\vec{\rho}_\alpha}-
i\frac{\hbar}{M_\alpha}\vec{q}_\alpha\cdot\vec{\nabla}_{\vec{\rho}_\alpha}
+\frac{Z_\alpha(Z_\beta+Z_\gamma) e^2}{4 \pi
\epsilon_0}\frac{1}{\rho_\alpha}\right]
C_{\eta_{q_\alpha}}(\zeta_{q_\alpha})=0,
\end{equation}
and
$$\eta_q\equiv Z_\alpha(Z_\beta+Z_\gamma)\left[\frac{e^2}{4 \pi \epsilon_0
\hbar}\right]\frac{M_\alpha}{\hbar q_\alpha}.$$

While equation (\ref{eq:exact}) constitutes a rigorous solution it is not a valid
solution because it does not match smoothly with the result asserted by Redmond.
(cf., equation (\ref{eq:redmond}) above.) To see this note that the asymptotic form
of (\ref{eq:exact}) in $\Omega_0$ would be given by
\begin{equation}
\Psi_{\vec{k}_\alpha,\vec{q}_\alpha}^{(\Omega_0)}
(\vec{r}_\alpha,\vec{\rho}_\alpha)=
e^{i(\vec{k}_\alpha\cdot\vec{r}_\alpha+\vec{q}_\alpha\cdot\vec{\rho}_\alpha)}
e^{-i\eta_{k_\alpha}\ln\zeta_{k_\alpha}}e^{-i\eta_{q_\alpha}\ln\zeta_{q_\alpha}}
\end{equation}
Clearly, the \emph{two} logarithmic phases present in this solution can not match
smoothly with the \emph{three} present in the Redmond solution. Hence
(\ref{eq:exact}) does not constitute a valid solution in the asymptotic region
$\Omega_0$. In addition, note that as a general wavefunction in the region
$\Omega_0$, (\ref{eq:exact}) may not be a good choice simply because when the two
charges $Z_\beta$ and $Z_\gamma$ are of equal and opposite charge, as in the paradigm
case of electron scattering from hydrogen, the solution in $\vec{\rho_\alpha}$
reduces to that of a free particle. i.e.,
\begin{equation}
\Psi_{\vec{k}_\alpha,\vec{q}_\alpha}^{(\Omega_\alpha,Z_\beta=-Z_\gamma)}
(\vec{r}_\alpha,\vec{\rho}_\alpha)=
e^{i(\vec{k}_\alpha\cdot\vec{r}_\alpha+\vec{q}_\nu\cdot\vec{\rho}_\nu)}
C_{\eta_{k_\alpha}}(\zeta_{k_\alpha}).
\end{equation}

These reasons make it clear that in order to extend the 3C
wavefunction into the the asymptotic region, $\Omega_\alpha$ (or
into the interior regions for that matter) a different approach
is warranted. The current approach, refered to here as ``kinematic
coupling,'' involves the introduction of an \emph{exact}
interaction potential given by:
\begin{equation}
V_I^\alpha(\vec{r}_\alpha,\vec{\rho}_\alpha) \equiv
\frac{e^2}{4\pi\epsilon_0} \left(\frac{Z_\alpha Z_\gamma}
{\left|\vec{r}_{\alpha}-\frac{\mu_\alpha}{m_\gamma}
\vec{\rho}_\alpha \right|}+ \frac{Z_\alpha Z_\beta}
{\left|\vec{r}_{\alpha}+\frac{\mu_\alpha}{m_\beta}
\vec{\rho}_\alpha\right|}-\frac{Z_\alpha(Z_\beta+Z_\gamma)}{\rho_\alpha}
\right),
\end{equation}
Hence the three-body Schr\"odinger equation takes the \emph{exact} form:
\begin{equation}\label{eq:myschrodinger}
\left[-\frac{\hbar^2}{2\mu_\alpha}\vec{\nabla}^2_{\vec{r}_\alpha}
+\frac{Z_\beta Z_\gamma e^2}{4 \pi \epsilon_0}\frac{1}{r_\alpha}
-\frac{\hbar^2}{2M_\alpha}\vec{\nabla}^2_{\vec{\rho}_\alpha}
+\frac{Z_\alpha(Z_\beta+Z_\gamma) e^2}{4 \pi \epsilon_0}\frac{1}{\rho_\alpha}
+V_I^\alpha(\vec{r}_\alpha,\vec{\rho}_\alpha)\right]
\Psi_{\vec{k}_\alpha,\vec{q}_\alpha}
(\vec{r}_\alpha,\vec{\rho}_\alpha)= {\mathcal{E}}_t
\Psi_{\vec{k}_\alpha,\vec{q}_\alpha} (\vec{r}_\alpha,\vec{\rho}_\alpha).
\end{equation}

To remain completely general, one then asserts that the total center of mass energy
may be written in the form
\begin{equation}\label{eq:myenergy}
{\mathcal{E}}_t={\mathcal{E}}_{\mu_\alpha}+{\mathcal{E}}_{M_\alpha}+{\mathcal{E}}_I,
\end{equation}
where ${\mathcal{E}}_{\mu_\alpha}\equiv\frac{\hbar@2k^2_\alpha}{2\mu_\alpha}>0$ and
${\mathcal{E}}_{M_\alpha}\equiv\frac{\hbar^2q^2_\alpha}{2 M_\alpha}>0$ are the
energies associated with the continuum state, two-body clusters, $\mu_\alpha$ and
$M_\alpha$ respectively, and the term ${\mathcal{E}}_I$ accounts for the remaining
energy of the three-body interaction. Though the definition (\ref{eq:myenergy}) is
nonstandard, it reflects the fact that because the three-body potential must remain
inseparable in a completely general solution, so too must the total energy.

Due to the fact that the inseparable portion of the three-body Coulomb potential is
now contained within the interaction potential, one can assume that the three-body
wavefunction takes the form,
\begin{equation}\label{eq:mypsi}
\Psi_{\vec{k}_\alpha,\vec{q}_\alpha}
(\vec{r}_\alpha,\vec{\rho}_\alpha)=
e^{i(\vec{k}_\alpha\cdot\vec{r}_\alpha+\vec{q}_\alpha\cdot\vec{\rho}_\alpha)}
C_{\eta_{k_\alpha}}(\zeta_{k_\alpha}) C_{\eta_{q_\alpha}}(\zeta_{q_\alpha})
\chi_{\vec{k}_{\alpha},\vec{q}_{\alpha}}(\vec{r}_{\alpha},\vec{\rho}_{\alpha}),
\end{equation}
where
$\chi_{\vec{k}_{\alpha},\vec{q}_{\alpha}}(\vec{r}_{\alpha},\vec{\rho}_{\alpha})$ is
an \emph{unknown} function, and $C_{\eta_{k_\alpha}}(\zeta_{k_\alpha})$ and
$C_{\eta_{q_\alpha}}(\zeta_{q_\alpha})$ are the \emph{known} Coulomb waves defined in
equations (\ref{eq:cka}) and (\ref{eq:cqa}) respectively. That is, the ansatz
(\ref{eq:mypsi}) incorporates the exact solution in the region $\Omega_\alpha$ (cf.,
equation (\ref{eq:exact})) explicitly.

Substitution of the form (\ref{eq:mypsi}) into the three-body Schr\"odinger equation
(\ref{eq:myschrodinger}) yields the following result:
\begin{eqnarray}\label{eq:chi}
&C_{\eta_{q_\alpha}}(\zeta_{q_\alpha})
\chi_{\vec{k}_{\alpha},\vec{q}_{\alpha}}(\vec{r}_{\alpha},\vec{\rho}_{\alpha})
\left(-\frac{\hbar^2}{2\mu_\alpha}\vec{\nabla}^2_{\vec{r}_\alpha}-
i\frac{\hbar}{\mu_\alpha}\vec{k}_\alpha\cdot\vec{\nabla}_{\vec{r}_\alpha}
+\frac{Z_\beta Z_\gamma
e^2}{4\pi\epsilon_0}\frac{1}{r_\alpha}\right)C_{\eta_{k_\alpha}}(\zeta_{k_\alpha})&\nonumber\\
&+C_{\eta_{k_\alpha}}(\zeta_{k_\alpha})
\chi_{\vec{k}_{\alpha},\vec{q}_{\alpha}}(\vec{r}_{\alpha},\vec{\rho}_{\alpha})
\left(-\frac{\hbar^2}{2M_\alpha}\vec{\nabla}^2_{\vec{\rho}_\alpha}-
i\frac{\hbar}{M_\alpha}\vec{q}_\alpha\cdot\vec{\nabla}_{\vec{\rho}_\alpha}
+\frac{Z_\alpha(Z_\beta+Z_\gamma)e^2}{4\pi\epsilon_0}\frac{1}{\rho_\alpha}\right)C_{\eta_{q_\alpha}}(\zeta_{q_\alpha})&\nonumber\\
&-C_{\eta_{k_\alpha}}(\zeta_{k_\alpha}) C_{\eta_{q_\alpha}}(\zeta_{q_\alpha})
\left(-\frac{\hbar^2}{2\mu_\alpha}\vec{\nabla}^2_{\vec{r}_\alpha}-
\frac{\hbar^2}{2M_\alpha}\vec{\nabla}^2_{\vec{\rho}_\alpha}-
i\frac{\hbar}{\mu_\alpha}\vec{k}_\alpha\cdot\vec{\nabla}_{\vec{r}_\alpha}-
i\frac{\hbar}{M_\alpha}\vec{q}_\alpha\cdot\vec{\nabla}_{\vec{\rho}_\alpha}\right)
\chi_{\vec{k}_{\alpha},\vec{q}_{\alpha}}(\vec{r}_{\alpha},\vec{\rho}_{\alpha})&\\
&-\frac{\hbar^2}{\mu_\alpha}C_{\eta_{q_\alpha}}(\zeta_{q_\alpha})\vec{\nabla}_{\vec{r}_\alpha}
C_{\eta_{k_\alpha}}(\zeta_{k_\alpha})\cdot
\vec{\nabla}_{\vec{r}_\alpha}\chi_{\vec{k}_{\alpha},\vec{q}_{\alpha}}(\vec{r}_{\alpha},\vec{\rho}_{\alpha})-
\frac{\hbar^2}{M_\alpha}C_{\eta_{k_\alpha}}(\zeta_{k_\alpha})\vec{\nabla}_{\vec{r}_\alpha} C_{\eta_{q_\alpha}}(\zeta_{q_\alpha})\cdot
\vec{\nabla}_{\vec{\rho}_\alpha}\chi_{\vec{k}_{\alpha},\vec{q}_{\alpha}}(\vec{r}_{\alpha},\vec{\rho}_{\alpha})&\nonumber\\
&+V_\alpha^I(\vec{r}_\alpha,\vec{\rho}_\alpha)C_{\eta_{k_\alpha}}(\zeta_{k_\alpha})C_{\eta_{q_\alpha}}(\zeta_{q_\alpha})
\chi_{\vec{k}_{\alpha},\vec{q}_{\alpha}}(\vec{r}_{\alpha},\vec{\rho}_{\alpha})=
{\mathcal{E}}_I C_{\eta_{k_\alpha}}(\zeta_{k_\alpha})
C_{\eta_{q_\alpha}}(\zeta_{q_\alpha})
\chi_{\vec{k}_{\alpha},\vec{q}_{\alpha}}(\vec{r}_{\alpha},\vec{\rho}_{\alpha}),&\nonumber
\end{eqnarray}
where the exponential terms have been cancelled from both sides of the equation and
the following vector identities have been employed:
$$
\vec{\nabla}(AB)=B\vec{\nabla}A+A\vec{\nabla}B,
$$
and
$$
\vec{\nabla}^2(AB)=B\vec{\nabla}^2{A}+A\vec{\nabla}^2{B}+2\vec{\nabla}A\cdot\vec{\nabla}B.
$$

Using equations (\ref{eq:cka}) and (\ref{eq:cqa}) in conjunction
with the definition (\ref{eq:myenergy}), one finds that the first
two lines in this unwieldy expression are identically zero. Then
after dividing by the product
$C_{\eta_{k_\alpha}}(\zeta_{k_\alpha})C_{\eta_{q_\alpha}}(\zeta_{q_\alpha})$,
the following equation for the unknown function,
$\chi_{\vec{k}_{\alpha},\vec{q}_{\alpha}}(\vec{r}_{\alpha},\vec{\rho}_{\alpha})$
is derived:
\begin{equation}\label{eq:chiwave}
\left[-\frac{\hbar^2}{2\mu_\alpha}\vec{\nabla}^2_{\vec{r}_\alpha}-
i\frac{\hbar}{\mu_\alpha}\vec{K}(\vec{r}_\alpha)\cdot\vec{\nabla}_{\vec{r}_\alpha}-
\frac{\hbar^2}{2M_\alpha}\vec{\nabla}^2_{\vec{\rho}_\alpha}-
i\frac{\hbar}{M_\alpha}\vec{Q}(\vec{\rho}_\alpha)
\cdot\vec{\nabla}_{\vec{\rho}_\alpha}+
V_\alpha^I(\vec{r}_\alpha,\vec{\rho}_\alpha)\right]
\chi_{\vec{k}_{\alpha},\vec{q}_{\alpha}}(\vec{r}_{\alpha},\vec{\rho}_{\alpha})=
{\mathcal{E}}_I\chi_{\vec{k}_{\alpha},\vec{q}_{\alpha}}(\vec{r}_{\alpha},\vec{\rho}_{\alpha}),
\end{equation}
is found, where
\begin{subequations}\label{eq:momenta}
\begin{eqnarray}
\vec{K}(\vec{r}_\alpha)&\equiv& \vec{k}_\alpha-
i\frac{\vec{\nabla}_{\vec{r}_\alpha}
C_{\eta_{k_\alpha}}(\zeta_{k_\alpha})}{C_{\eta_{k_\alpha}}(\zeta_{k_\alpha})},\\
\vec{Q}(\vec{\rho}_\alpha)&\equiv&\vec{q}_\alpha-i\frac{\vec{\nabla}_{\vec{\rho}_\alpha}C_{\eta_{q_\alpha}}(\zeta_{q_\alpha})}
{C_{\eta_{q_\alpha}}(\zeta_{q_\alpha})},
\end{eqnarray}
\end{subequations}
are the proposed position-dependent local momenta for this kinematic coupling model.

The principle difference in this development is that the local momenta arise purely
from the physical structure of the inseparable three-body Schr\"odinger equation, and
that their mathematical form implies the existence of position dependent momenta.
Furthermore, in the current development, the local momenta are predicted in a
symmetric fashion so that both ``legs'' of the three-body interaction experience
distortions which depend upon the congugate coordinates. That is, $\vec{r}_\alpha$ is
congugate to $\vec{k}_\alpha$ and $\vec{\rho}_\alpha$ is congugate to
$\vec{q}_\alpha$. This means that the distortion that two of the particles experience
does not depend explicitly upon the distance to the third particle as in \cite{AM},
but is rather wholly attributable to a genuine three-body effect wherein a very
intuitive description of the three-body scattering event arises:
\begin{quote}
The two reduced mass clusters $\mu_\alpha$ and $M_\alpha$, initially described by the
two-body waves, $C_{\eta_{k_\alpha}}(\zeta_{k_\alpha})$ and
$C_{\eta_{q_\alpha}}(\zeta_{q_\alpha})$ respectively, scatter off of the interaction
potential $V_I(\vec{r}_{\alpha},\vec{\rho}_{\alpha})$ and acquire a local momenta.
\end{quote}
The subsequent motion of these clusters is then of course dictated by the function
$\chi_{\vec{k}_{\alpha},\vec{q}_{\alpha}}(\vec{r}_{\alpha},\vec{\rho}_{\alpha})$
which must satisfy (\ref{eq:chiwave}).

While one may argue that solving equation (\ref{eq:chiwave}) is a greater task than
solving the original Schhr\"odinger equation, a solution that is consistent with the
3C wavefunction in the region $\Omega_0$ may be established. The details of this
solution are not important to the present discussion and are relegated to the
appendices. (See appendix \ref{app:kc5c}.) What will be of great importance is the
nature of the predicted local momenta, (\ref{eq:momenta}).

\section{A Generalized Local Momentum}
\label{sec:general}

The \emph{local momenta} introduced in this \emph{kinematically coupled} model depend
upon the \emph{congugate} coordinate. Because of this variations in the momenta will
generally contribute to the solutions. To understand how and where these
contributions will be important, a generalized ``local distortion'' term,
$\vec{D}_{\vec{p}}(\vec{r})$ is introduced, where $\vec{p}$ and $\vec{r}$ may be
either $\vec{k}_\nu$ or $\vec{q}_\nu$ and $\vec{r}_\nu$ or $\vec{\rho}_\nu$,
respectively. Hence a general position-dependent momentum may be defined as follows:
\begin{equation}
\vec{P}(\vec{r})=\vec{p}-\vec{D}_{\vec{p}}(\vec{r}),
\end{equation}
where the exact form of the local distortion is given by\footnote{See Appendix
\ref{app:1f1} for details concerning the derivation of this form.}:
\begin{subequations}
\begin{eqnarray}
\label{eq:distdefined}
\vec{D}_{\vec{p}}(\vec{r})&\equiv&i\vec{\nabla}_{r} \ln C_{\eta_{p}}= i\eta_p p
\frac{_{1}F_{1}[(1+i\eta_{p}),2,-i\zeta_{p}]} {_{1}F_{1}(i\eta_{p},1,-i\zeta_{p})}
(\hat{p}+\hat{r})\\
\label{eq:distderived}
\vec{D}_{\vec{p}}(\vec{r})&=&\left\{
\begin{tabular}{cc} $i\eta_{p} p (\hat{p}+\hat{r})$ & $,r=0$\\
$\frac{\eta_p}{r}
\left[1-e^{-i(\zeta_{p}+2\delta_{\eta_{p}}(\zeta_{p}))}\right]
\frac{(\hat{p}+\hat{r})}{\hat{p}\cdot(\hat{p}+\hat{r})}$ & $,r>0$
\end{tabular}.
\right.
\end{eqnarray}
\end{subequations}

Here
\begin{equation}\label{eq:etap}
\eta_p=Z \left(\frac{e^2}{4\pi\epsilon_0\hbar}\right)\frac{\mu}{\hbar p}=
Z\left(\frac{e^2}{4\pi\epsilon_0\hbar}\right)
\sqrt{\frac{\mu}{2E}},
\end{equation}
$C_{\eta_p}(\zeta_p)$ satisfies
\begin{equation}
\left(-\frac{\hbar^2}{2\mu}\vec{\nabla}_{\vec{r}}^2
-i\frac{\hbar}{\mu}\vec{p}\cdot\vec{\nabla}_{\vec{r}}
+\frac{Ze^2}{4\pi\epsilon_0}\frac{1}{r}\right)C_{\eta_p}(\zeta_p)=0,
\end{equation}
and $\delta_{\eta_{p}}(\zeta_{p})$ is a \emph{real}-valued,
position dependent phase, defined by
\begin{equation}
\tan\left[\delta_{\eta_{p}}(\zeta_{p})\right]=
\frac{\Im[_{1}F_{1}(i\eta_{p},1,-i\zeta_{p})]}
{\Re[_{1}F_{1}(i\eta_{p},1,-i\zeta_{p})]}.
\end{equation}

While the general form (\ref{eq:distdefined}) has been used by many researchers, the form (\ref{eq:distderived})
is unique and has been introduced to help illucidate the physical significance of the local momentum.
e.g., one may write: ($r\not=0$)
\begin{equation}\label{eq:distortion}
\vec{D}_{\vec{p}}(\vec{r})=Z\left(\frac{e^2}{4\pi\epsilon_0\hbar}\right)\sqrt{\frac{\mu}{2E}}
\frac{a_{\eta_p}^*(\zeta_p)}{r}(\hat{p}+\hat{r}),
\end{equation}
where $Z$ and $\mu$ are the relevant \emph{product charge} and
\emph{reduced mass} of the pair described by $C_{\eta_p}(\zeta_p)$, and
\begin{equation}\label{eq:mya}
a_{\eta_p}^*(\zeta_p)\equiv\frac{\left[1-e^{i\left(\zeta_p+2\delta_{\eta_p}(\zeta_p)\right)}\right]}
{(1+\hat{p}\cdot\hat{r})}.
\end{equation}

The form (\ref{eq:distortion}) shows that the local momentum is essentially a
$\frac{1}{r}$ damped-oscillator function, and that it is analytic everywhere except
possibly at $r$ equal to zero. Interestingly, it can also be seen that the local
momentum has equal radiation (in the $\hat{r}$ direction) and induction (in the
$\hat{p}$ direction) components. This is indicative of the possibility of a tensor
force in a Classical treatment or of a nonconserved current density in a Quantum
treatment of the three-body interaction. Perhaps more importantly, one sees that the
local distortion depends upon not only the radial separation of the constituent
particles, but also upon both the relative energy and reduced mass of the system.

\begin{figure}
\begin{center}
$\quad\quad\quad$\fbox{$\text{${\mathcal{E}}$ Dependence of the
Local Distortion}$}\\ \psfrag{r}[bc]{$r\ (a_0)$}
\psfrag{D}[bc]{$\Re D_{\vec{p}}(\vec{r})\ (\frac{1}{p})$}
\psfrag{Re}{$\Re$} \psfrag{Im}{$\Im$} \psfrag{eV}{$\mathrm{eV}$}
\resizebox{3in}{!}{\includegraphics{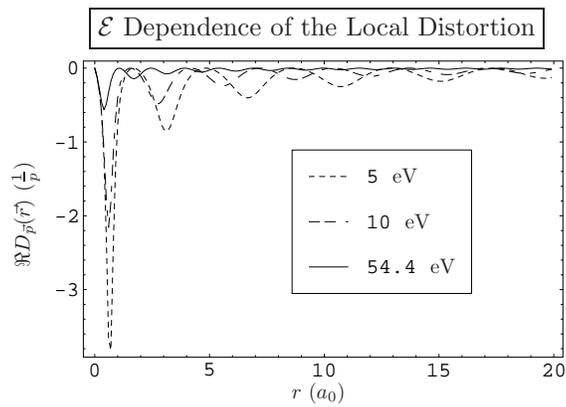}}
\end{center}
\caption{The real part of the local distortion experienced by an electron-proton
continuum pair with relative energies as indicated and a scattering angle of
$\theta=4^\circ$.}
\label{fig:variousenergies}
\end{figure}

In Figure \ref{fig:variousenergies} the \emph{real} part of the distortion experienced
by a electron-proton continuum pair has been plotted with various relative energies
and an arbitrary scattering angle of $\theta=0$. One sees immediately that the
distortion ``falls off'' very quickly with increasing radial separation and even more
dramatically with increased relative energy.

\begin{figure}
\begin{center}
$\quad\quad\quad$\fbox{$\text{$\mu$ and $Z$ Dependence of the
Local
Distortion}$}\\
 \psfrag{r}[bc]{$r\ (a_0)$}
\psfrag{D}[bc]{$\Re D_{\vec{p}}(\vec{r})\ (\frac{1}{p})$}
\psfrag{ee}{$e-\bar{e}$} \psfrag{ep}{$e-p$}
\resizebox{3in}{!}{\includegraphics{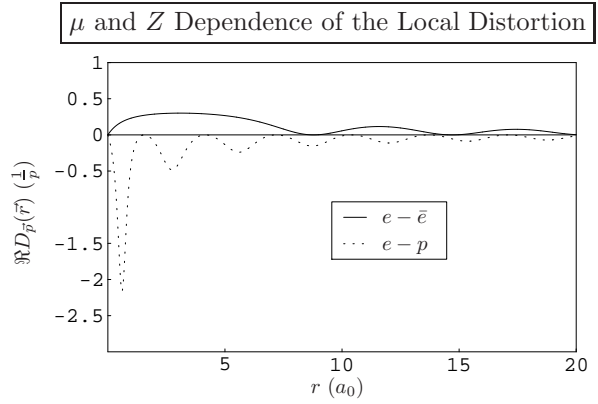}}
\end{center}
\caption{The real part of the local distortion experienced by both electron-electron
and electron-proton continuum pairs, with values as in Fig. \ref{fig:variousenergies}.
Note that the sign of the distortion is different for the attractive $(e^-,p^+)$ and
repulsive $(e^-,e^-)$ cases.}
\label{fig:variousmasses}
\end{figure}

To see how the distortion depends upon the reduced mass of the system, observe that
in Figure \ref{fig:variousmasses}, the \emph{real} part of the distortion experienced
by an electron-electron pair and a electron-proton pair have been plotted. Notice
that the distortion increases accordingly with increased mass, and that the sign of
the distortion is different for the cases of attraction and repulsion.

This very interesting point will be discussed in greater detail
below, but for now note that in addition to the important
physical properties of the local distortion, there are also
important mathematical properties. The most important of these is
that because the local distortion is essentially an oscillatory
function, the existence of turning and stationary (i.e., maxima
and minima) points may yield much insight.

\begin{figure}
\begin{center}
$\quad\quad\quad$\fbox{$\text{Stationary Points of the Local
Distortion}$}\\ \psfrag{r}[bc]{$r\ (a_0)$}
\psfrag{D}[bc]{$D_{\vec{p}}(\vec{r})\ (\frac{1}{p})$}
\psfrag{Re}{$\Re$} \psfrag{Im}{$\Im$}
\resizebox{3in}{!}{\includegraphics{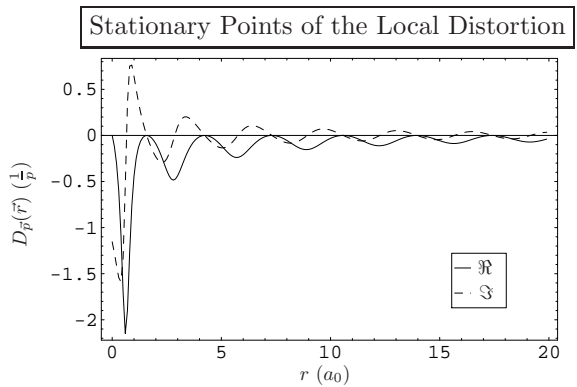}}
\end{center}
\caption{The local distortion experienced by an electron-proton continuum pair
with a relative energy of $10\ \mathrm{eV}$, and a scattering angle of $\theta=16^\circ$.
Note that when the zeroes of \emph{real} and \emph{imaginary} part of the distortion
coincide, the distortion contribution will be identically zero.}
\label{fig:stationarypoints}
\end{figure}

\begin{figure}
\begin{center}
\fbox{$\text{Topology of the Local Distortion}$}\\
\psfrag{C}[bc]{\rotatebox{90}{$\Re D_{\vec{p}}(\vec{r})\
(\frac{1}{p})$}} \psfrag{r}{$r\ (a_0)$} \psfrag{theta}{$\theta$}
\psfrag{Dep}{$\Re D$}
\resizebox{3in}{!}{\includegraphics{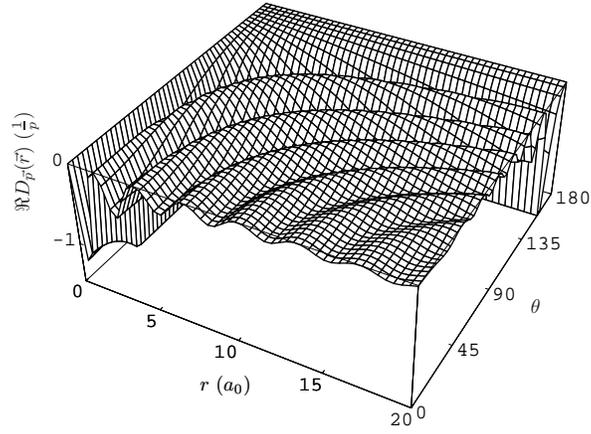}}\\
(a)\\
\vspace{0.2in}
\resizebox{3in}{!}{\includegraphics{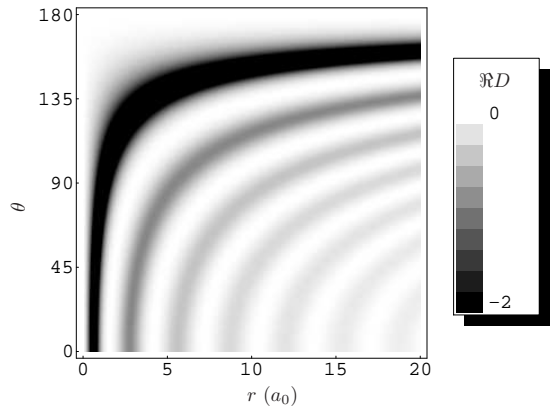}}\\
(b)
\end{center}
\caption{Shown here are (a) three-dimensional and (b) contour plots of the \emph{real}
part of the local distortion experienced by an electron-proton continuum pair with a
relative energy of $10\ \mathrm{eV}$.}
\label{fig:visual}
\end{figure}

In Figure \ref{fig:stationarypoints} one sees that the stationary
points\footnote{That is, the points where the zeroes of the real and imaginary parts
of the local distortion coincide.} occur quite often, so that the local distortion
will very often be identically zero! Moreover, these zeroes will be very dense on a
macroscopic scale and so may greatly alter the topology of a scattering event. This
can be seen even more dramatically in Figure \ref{fig:visual}, where
three-dimensional and contour plots of the local distortion have been shown.

\begin{figure}
\begin{center}
$\quad\quad\quad$\fbox{$\text{Local Distortion Phase Effects}$}
\psfrag{r}[bc]{$r\ (a_0)$}\\
\psfrag{delta}[bc]{$\delta_{\eta_p}(\zeta_p)$}
\psfrag{theta}{$\theta=20^\circ$} \psfrag{eV}{$\mathrm{eV}$}
\resizebox{3in}{!}{\includegraphics{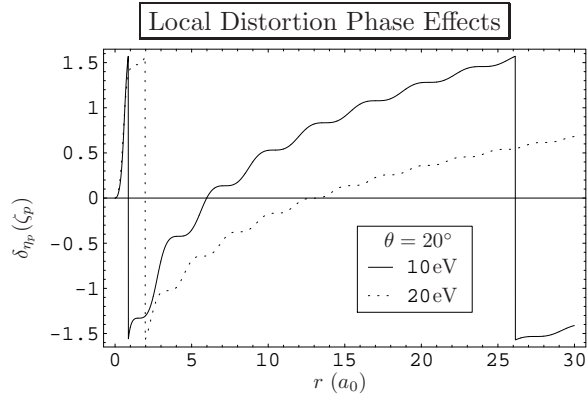}}
\end{center}
\caption{Shown here is the position dependent-phase, $\delta_{\eta_p}(\zeta_p)$.
Note that it is very nearly constant over one log-cycle and that it is nearly
independent of the relative energy of the electron-proton continuum pair.}
\label{fig:delta}
\end{figure}

Though there is no analytic form for determining the zeroes of the confluent
hypergeometric function, the stationary points of the local distortion may be found
by finding the nonzero roots to (\ref{eq:mya}). Hence one ends up solving the following
transcendental equation:
\begin{equation}\label{eq:myconic}
\frac{1}{r}=\frac{p}{2 [n\pi-\delta_p(\vec{r})]}(1-\cos\theta),
\end{equation}
where $n$ is an integer. Note that if $\delta_p(\vec{r})$ were a constant,
(\ref{eq:myconic}) would be the equation of a conic section with an eccentricity of
1 - parabolas. Hence one sees that the stationary points of the local distortion
lie (very nearly) along the classically forbidden ``trajectories'' for particles with
positive energy! While this may seem to be no more than coincidence, $\delta_p(\vec{r})$
is in fact very nearly constant between the stationary points (see Fig. \ref{fig:delta}),
and one may infer that these ``zero-distortion-trajectories'' are in fact the dominant
contributions in a path-integral approach to the quantum, three-body scattering problem.
Indeed, from this point of view, one may argue that they must be the dominant
contributions for light-ion--atom processes in the asymptotic regions, where great
success has been achieved while ignoring local distortion effects. What is not clear
however is how these local effects will contribute in the so called ``interior regions''
or for heavy-ion--atom processes.

\section{Kinematic Rearrangement}
\label{sec:rearrange}

Above it was noted that the local distortion had different signs for the attractive and
repulsive cases. The real importance of this observation is that if one imagines a
three-body system composed of a tightly bound continuum state pair, with an initial
momentum $\vec{p}$, and a third, mutually interacting particle, then upon break-up the
continuum pair will acquire a local momentum $\vec{P}(\vec{r})$. Hence the continuum
pair will experience a momentum change (or \emph{impulse}) given by
\begin{equation}
\vec{I}(\vec{r})\equiv
\Re\vec{P}(\vec{r})-\vec{p}=-\Re \vec{D}_{\vec{p}}(\vec{r}).
\end{equation}
Evidently one finds, (for $r\not=0$)
\begin{subequations}\label{eq:impulse}
\begin{eqnarray}
\vec{I}_{\mathrm{attractive}}&=&+|Z|\left(\frac{e^2}{4\pi\epsilon_0\hbar}\right)
\sqrt{\frac{\mu}{2 E}}
\frac{\left[1-\cos(\zeta_{p}+2\delta_{\eta_{p}}){}\right]}
{\hat{p}\cdot(\hat{p}+\hat{r})}\frac{1}{r}(\hat{p}+\hat{r})\\
\vec{I}_{\mathrm{repulsive}}&=&-|Z|\left(\frac{e^2}{4\pi\epsilon_0\hbar}\right)
\sqrt{\frac{\mu}{2 E}}
\frac{\left[1-\cos(\zeta_{p}+2\delta_{\eta_{p}}){}\right]}
{\hat{p}\cdot(\hat{p}+\hat{r})}\frac{1}{r}(\hat{p}+\hat{r}).
\end{eqnarray}
\end{subequations}

The equations (\ref{eq:impulse}) imply that the two particles that are initially
attracted to one another, within the continuum pair described by $C_{\eta_p}(\zeta_p)$,
will experience an impulse that will tend to increase their radial separation upon
breakup and conversely, that two particles that are initially repelled from one another
will experience an impulse that will tend to decrease their radial separation. In other words,
at small values of the radial separation, opposite(same) sign-charged particles
will tend to repel(attract) one another due to the local distortion effects!

\begin{figure}
\begin{center}
\fbox{$\text{Local Distortion Effects for Various Continuum
Pairs}$} \psfrag{r}[c]{$r\ (a_0)$} \psfrag{dist}[c]{$\Re
D_{\vec{p}}(\vec{r})\ (1/p)$}
$$
\begin{array}{cc}
{\psfrag{title}[bc]{(a)
\fbox{$(e-\bar{e})$}}\psfrag{eV}[c]{$\theta=16^\circ,E=10\
\mathrm{eV}$} \resizebox{2in}{!}{\includegraphics[bb=0 350 285
540]{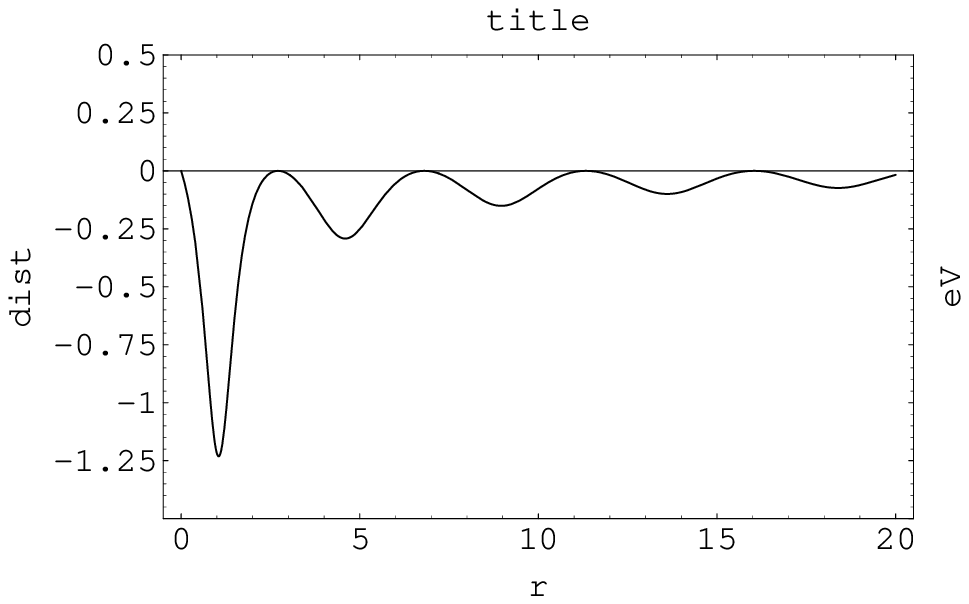}}} & {\psfrag{title}[bc]{(b)
\fbox{$(e-e)$}} \psfrag{eV}[c]{$\theta=16^\circ,E=10\
\mathrm{eV}$} \resizebox{2in}{!}{\includegraphics[bb=0 350 285
540]
{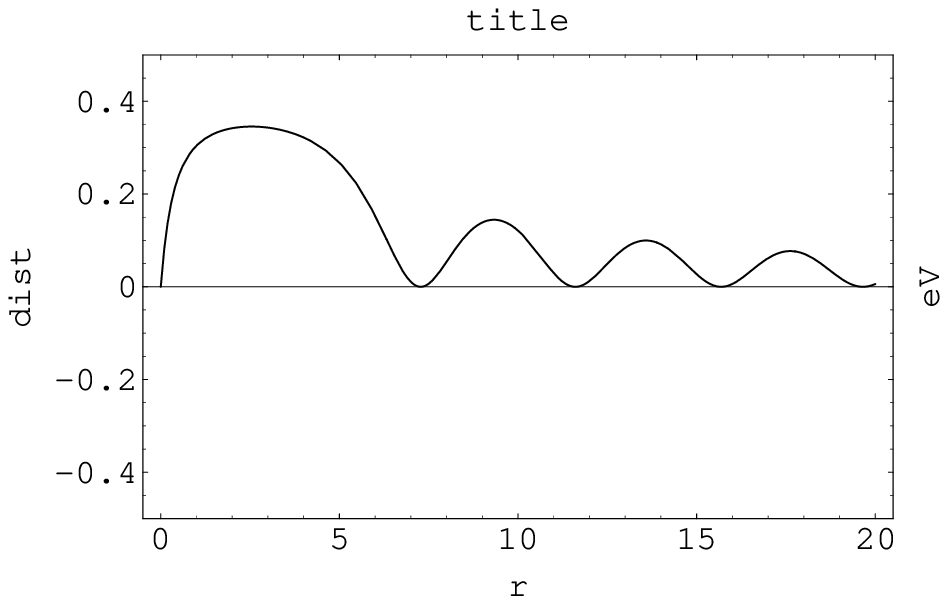}}} \\
{\psfrag{title}[bc]{(c)
\fbox{$(e-p)$}}\psfrag{eV}[c]{$\theta=16^\circ,E=10\
\mathrm{eV}$} \resizebox{2in}{!}{\includegraphics[bb=0 350 285
540]{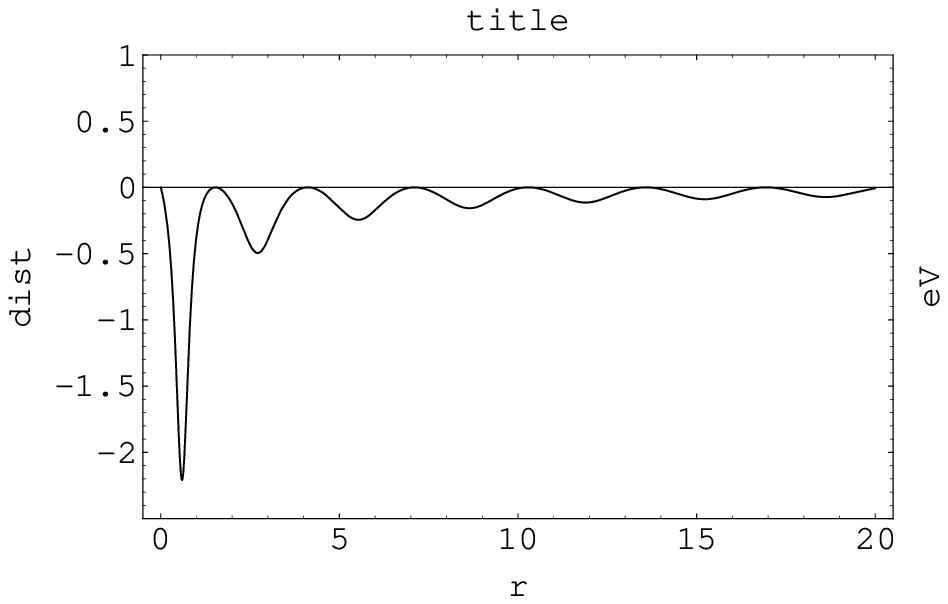}}} & {\psfrag{title}[bc]{(d)
\fbox{$(\bar{e}-\bar{p})$}} \psfrag{eV}[c]{$\theta=16^\circ,E=10\
\mathrm{eV}$} \resizebox{2in}{!}{\includegraphics[bb=0 350 285
540]
{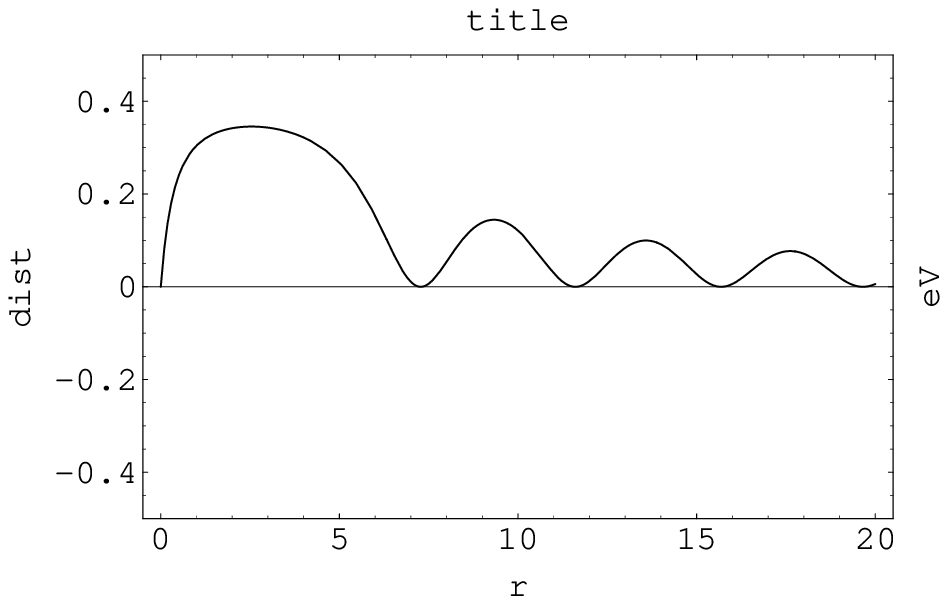}}} \\
{\psfrag{title}[bc]{(e) \fbox{$(p-\bar{p})$}}
\psfrag{eV}[c]{$\theta=16^\circ,E=50\ \mathrm{eV}$}
\resizebox{2in}{!}{\includegraphics[bb=0 350 285 540]{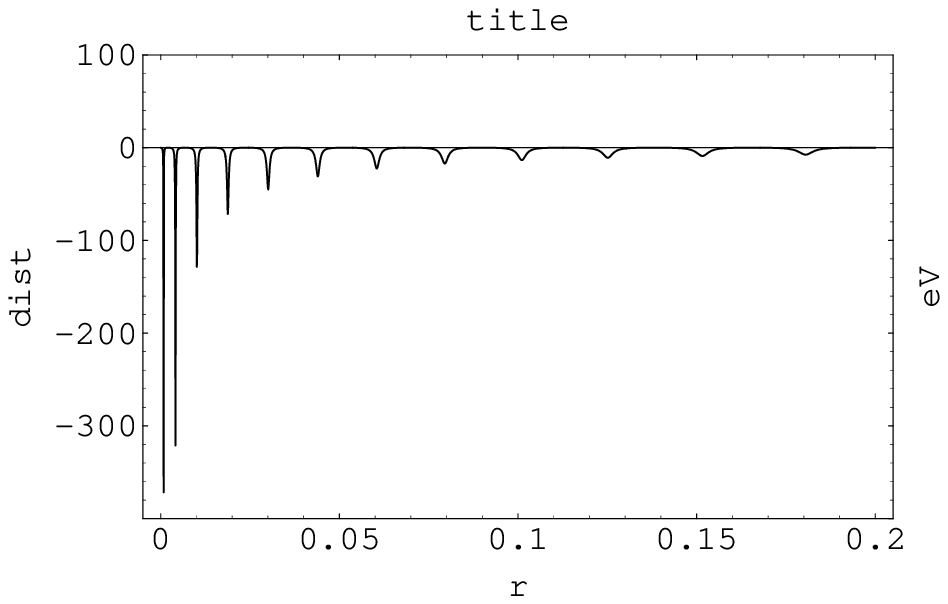}}}
& {\psfrag{title}[bc]{(f) \fbox{$(p-p)$}}
\psfrag{eV}[c]{$\theta=16^\circ,E=50\ \mathrm{eV}$}
\resizebox{2in}{!}{\includegraphics[bb=0 350 285 540]{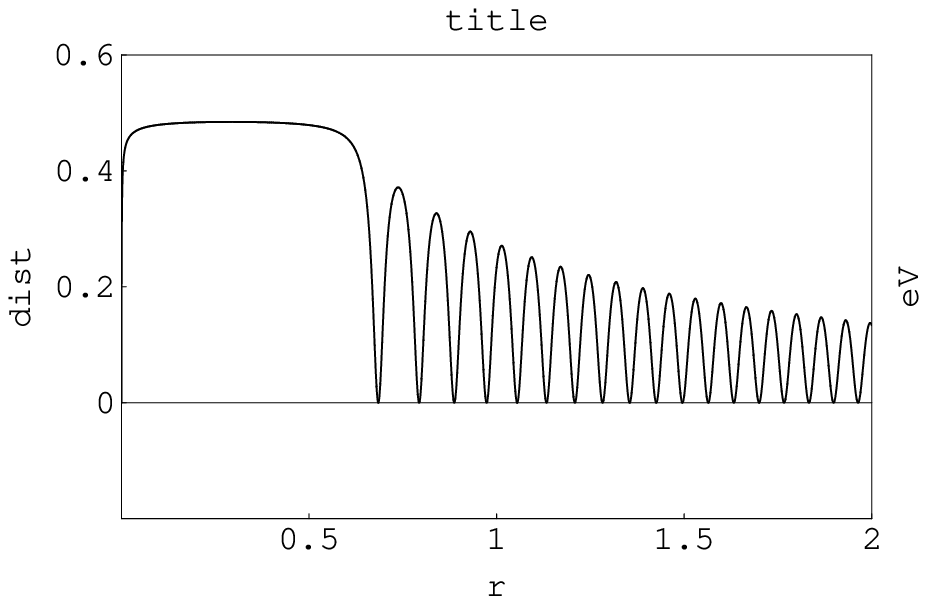}}}
\end{array}$$
\end{center}
\caption{Shown are the local distortion effects experienced by (a) electron-positron,
(b) electron-electron, (c) electron-antiproton , (d) positron-proton,
(e) proton-antiproton, and (f) proton-proton, continuum pairs with ${\mathcal{E}}=10\ \mathrm{eV}$
and $\theta=16^\circ$. While these effects become negligible
asymptotically, the effects become more pronounced in the interior regions; the
range has been shortened in (e) and (f) to show the dramatic change due to the increase
in the reduced mass of the pair}
\label{fig:variouspairs}
\end{figure}

To illustrate this point, the \emph{real} part of the distortion for several different
continuum pairs have been calculated and the results are shown in Figure
\ref{fig:variouspairs}. Observe that even though all of these distortion terms decrease
with increasing radial separation, the magnitude of the distortion in the interior
regions depends critically upon the system being studied. As was shown above in Figure
\ref{fig:variousmasses}, the distortion effects increases dramatically with an increase
in the reduced mass. Specifically, note that for the case of a proton--anti-proton
((e) if Fig. \ref{fig:variouspairs}) or proton-proton ((f) in Fig. \ref{fig:variouspairs}),
the range has been shortened accordingly, to show that the distortion in the reaction
zone is dramatically changed. Indeed, for the case of the proton--anti-proton, the
magnitude of the local distortion effect may be large enough and in a
direction such that a kinematic rearrangement of the particles may occur.
In addition, because of the dependence upon the inverse of the square-root of the
relative energy, these effects will be even more pronounced at lower relative
energies.

\begin{figure}
\begin{center}
\fbox{$\text{WYJ Dalitz Plot}\ \mathcal{E}_t=6.5\ \mathrm{eV},
\mathcal{P}_\mathrm{low}=\sqrt{\frac{31}{72}} $}\\
\resizebox{3in}{!}{\includegraphics[bb=0 0 479 223]{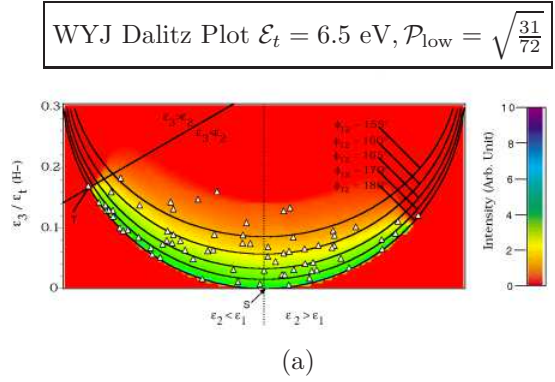}}\\
(a)\\
\vspace{0.2in} \fbox{$\text{WYJ Dalitz Plot}\ \mathcal{E}_t=7.5\
\mathrm{eV}, \mathcal{P}_\mathrm{high}=\sqrt{\frac{14}{47}} $}\\
\resizebox{3in}{!}{\includegraphics[bb=0 0 479 223]{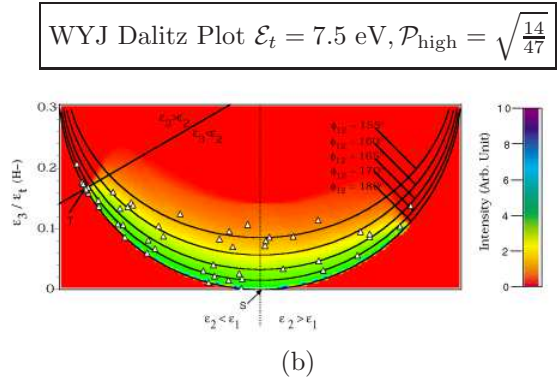}}\\
(b)
\end{center}
\caption{These original data, obtained directly from the authors, show that as the total
center-of-mass energy of the three-body system, ${\mathcal{E}}_t$, is increased, a proportionately
smaller number of particles undergoes kinematic rearrangement. Shown also are the predicted
probability ratios for this rearrangement for the (a) ${\mathcal{E}}_t=6.5\ \mathrm{eV}$ and
(b) ${\mathcal{E}}_t=7.5\ \mathrm{eV}$, triple-coincidence events.}
\label{fig:probability}
\end{figure}

\begin{figure}
\begin{center}
\fbox{$\text{Explanation of Three-Body Distortion Effects}$}\\
\psfrag{H3}{$(H_3^+)^*$} \psfrag{H2}{$H_2^{**}$}
\psfrag{Hip}{$H_i^+$} \psfrag{Hfp}{$H_f^+$} \psfrag{Hm}{$H^-$}
\psfrag{Hp}{$H^+$} \psfrag{p}{$\vec{p}$} \psfrag{r}{$\vec{r}$}
\psfrag{Pr}{$\vec{P}(\vec{r})$} \psfrag{CS}[bc]{Coulomb\ Saddle}
\psfrag{Low}[bc]{Low\ Energy} \psfrag{High}[bc]{High\ Energy}
\psfrag{a}{(a)} \psfrag{b}{(b)} \psfrag{c}{(c)} \psfrag{d}{(d)}
\resizebox{4in}{!}{\includegraphics[bb=5 5 600 785 ]{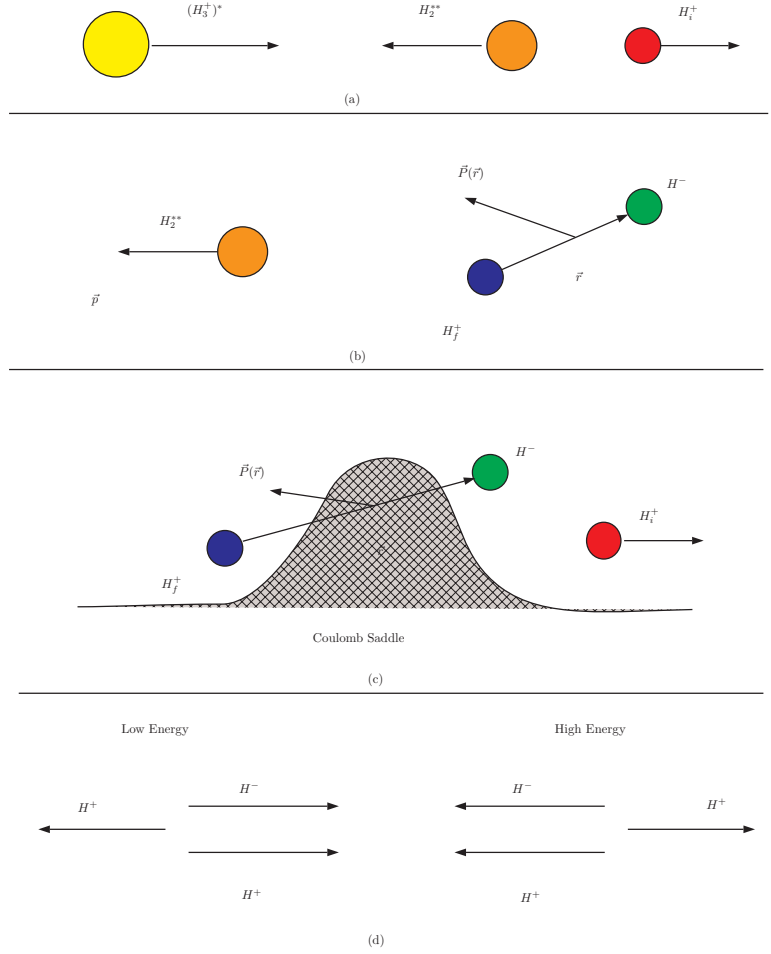}}
\end{center}
\caption{(a) During the breakup of the $H_3^+$ ion the $H_i^+$ is emitted in
the forward direction and the $H^{**}_2$ is emitted in the backward direction. (b) The
subsequent breakup of this ion occurs such that the $H^+_f$ and $H^-$ ions are formed, and
the $H^-$ is preferentially associated with $H^+_f$. (c) The continuum pair, ($H^+_{f},H^-$)
acquires a local momentum, $\vec{P}(\vec{r})$, pushing the $H^-$ over the Coulomb Saddle,
so that the $H^-$ will be associated with the $H^+_i$. (d) Because the two protons are in
fact indistinguishable, one will observe an asymmetry between
``low'' and ``high'' energy scattering events.}
\label{fig:explain}
\end{figure}

This kind of phenomena can be observed in the data obtained by Wiese \emph{et. al.},
\cite{Wiese} on the breakup of the excited ion $(H_3^+)^{*}$. This highly unstable ion
of hydrogen decays in a two step process, as follows:
$$(H_3^+)^{*}\longrightarrow H_2^{**}+H_i^+\ \rightarrow\ {H_f^+} +
{H^-} + {H_i^+}.$$ The data (see Figure \ref{fig:probability})
show a slight asymmetry with small increases in energy (from
$6.5\ \mathrm{eV}$ to $7.5\ \mathrm{eV}$) which can be accounted
for in a direct manner by considering the local distortion
effects herein derived. To see this note that during the nearly
co-linear breakup of the excited $H_3^+$ ion the
``initial''\footnote{``Initial'' and ``final'' are used here as
in \cite{Wiese} to indicate the order in which the particles were
detected during triple coincidence measurements.} proton $H_i^+$
is emitted in the forward direction and the doubly excited
$H^{**}_2$ ion is emitted in the backward direction. The
subsequent breakup of this ion occurs such that a ``final''
proton $H^+_f$ and the nearly equally weighted $H^-$ ion are
formed.\cite{Wiese}  At this point the $H^-$ is preferentially
associated with $H^+_f$. However, after the breakup, the
continuum pair ($H^+_{f},H^-$) will acquire a \emph{local
momentum}, $\vec{P}(\vec{r})$, which has a component in the
$\hat{r}$ direction. Hence this gained momentum will act to
increase the radial separation and will push the $H^-$ over the
Coulomb Saddle, so that it $H^-$ will be preferentially
associated with the ``initial'' proton, $H^+_i$. Now, because the
two protons are in fact indistinguishable, one will observe an
asymmetry between ``low'' and ``high'' energy scattering events.
To help visualize this process, a schematic diagram (see Fig.
\ref{fig:explain}.) has been developed.

\begin{figure}
\begin{center}
$\quad\quad$\fbox{$\text{Distortion Effects During}\ (H_3^-)^*\
\text{Breakup}$}\\ \psfrag{title}[bc]{} \psfrag{eV}{}
\psfrag{r}[bc]{$r\ (a_0)$} \psfrag{dist}[bc]{$\Re
D_{\vec{p}}(\vec{r})\ (\frac{1}{p})$}
\resizebox{3in}{!}{\includegraphics[bb=0 350 290
535]{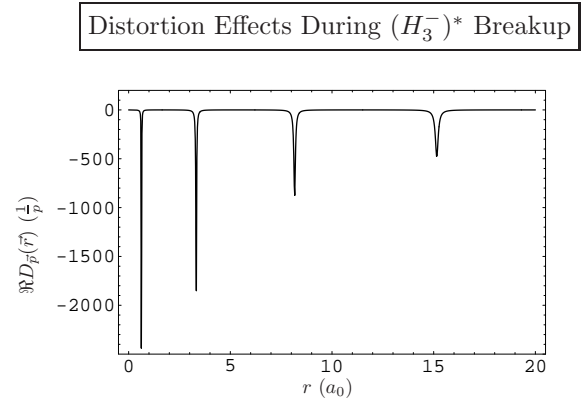}}
\end{center}
\caption{During the breakup of the $H_3^+$ ion, the magnitude of
the local distortion effects experienced by the ($H^+_{f},H^-$)
continuum pair while in the reaction zone are many orders of
magnitude larger than the effects experienced by either the
($e,\bar{e}$) or ($e,e$) continuum pairs. Here $\theta=176^\circ$
and ${\mathcal{E}}=1.5\ \mathrm{eV}$ for comparison with the
experiment.} \label{fig:hpbreakup}
\end{figure}

Further analysis shows that, as mentioned above, because the local
distortion depends upon $\frac{1}{\sqrt{E}}$, as the energy is
increased, a proportionately smaller number of particles will
experience an impulse that is large enough to alter their final
state distribution. Indeed, during the breakup of the $H_3^+$
ion, the magnitude of the local distortion effects experienced by
the ($H^+_{f},H^-$) continuum pair while in the reaction zone are
many orders of magnitude larger than the effects experienced by
an electron-proton continuum pair. (see Fig. \ref{fig:hpbreakup})
With this understanding, one may conclude in a straightforward
manner, that the observed asymmetry can be attributed to local
distortion effects of the three-body system. Moreover, the degree
of asymmetry may be predicted as follows:
\begin{equation}
\sqrt\frac{E_{\mathrm{high}}}{E_{\mathrm{low}}}\propto
\mathrm{\frac{amplitude\ of\ rearrangment\ at\ low\ energy}
{amplitude\ of\ rearrangement\ at\ high\ energy}}.
\end{equation}
Therefore, with the probabilistic interpretation of the wavefunction,
one may write,
\begin{equation}
\sqrt\frac{E_{\mathrm{high}}}{E_{\mathrm{low}}}
\equiv
\sqrt{\mathcal{\frac{P_{\mathrm{low}}}{P_{\mathrm{high}}}}}
\end{equation}
where $\mathcal{P}_{\mathrm{high}}$ and $\mathcal{P}_{\mathrm{low}}$ are
the probabilities of rearrangement for the high and low energy
states, respectively. That is, using the actual data (See Fig. \ref{fig:probability}.)
the ratio of the number of rearranged particles to the total number of
detected particles may be calculated. Doing this one obtains the following results:
$$\sqrt\frac{E_{\mathrm{high}}}{E_{\mathrm{low}}}=
\sqrt{\frac{7.5\ \mathrm{eV}}{6.5\ \mathrm{eV}}}=1.1$$
and
$${\mathcal{\frac{P_{\mathrm{low}}}{P_{\mathrm{high}}}}}=
\frac{\sqrt{\frac{31}{72}}}{\sqrt{\frac{14}{47}}}=1.2$$ The
absolute error between these two calculations is $\approx 8.7\%$
and illustrates that the generalized local momentum, herein
developed may in fact be the actual momentum of the ($H^-,H^+_f$)
continuum pair during breakup. Indeed, one may consider
revaluating the importance of the generalized local momenta.
Instead of treating it as a mere mathematical nicety, one may
place it on firm physical ground by viewing it is a formal gauge
condition for three-body interactions.

\section{Towards a Three-Body Gauge} \label{sec:gauge}

If one grants that the mathematical and computational evidence
gathered here, is sufficient to guide further experimental and
theoretical investigations by considering the generalized local
momentum as a formal three-body gauge condition, then one can
construct a  three-body gauge transformation. i.e., one may write:
\begin{equation}\label{eq:threebody}
\vec{p}\longrightarrow\
\vec{P}(\vec{r})=\vec{p}-\vec{D}_{\vec{P}}(\vec{r}),
\end{equation}
when working with three-body systems, in much the same way that one
would write,
\begin{equation}\label{eq:coulomb}
\vec{p}\longrightarrow\ \vec{P}(\vec{r})=\vec{p}-e\hbar\vec{A}(\vec{r}),
\end{equation}
when working with particles in an external magnetic field. (i.e., the Coulomb Gauge.)

The subtlety here is that while the transformation,
$$\vec{A}(\vec{r})\longrightarrow\vec{A}(\vec{r})+\vec{\nabla}_{\vec{r}}\Omega(\vec{r})$$
preserves the form of the magnetic field,
$\vec{B}(\vec{r})=\vec{\nabla}_{\vec{r}}\times\vec{A}(\vec{r})$,
and provides for the \emph{global gauge symmetry} of the
electromagnetic interaction, the condition (\ref{eq:threebody})
will contribute a \emph{real}-valued, position-dependent phase to
the relevant wavefunction.\footnote{It is interesting to note however that,
$\vec{\nabla}_{\vec{r}} \times \vec{D}_{\vec{p}}(\vec{r})=
i\vec{\nabla}_{\vec{r}} \times \vec{\nabla}_{\vec{r}} \ln
C_{\eta_p}(\zeta_p)\equiv\vec{0}.$} Thus any proposed three-body
interaction mechanism would require a \emph{local gauge
symmetry}.\footnote{For those readers that are unfamiliar with
the terms \emph{global} and \emph{local} in this context, note
that a global gauge transformation is one that preserves the
modulus of the appropriate wavefunction. eg.,
$\Psi(\vec{r})\rightarrow{}e^{-ia}\Psi(\vec{r})$. On the other
hand, a local gauge transformation generally does not and may take
the form,
$\Psi(\vec{r})\rightarrow{}e^{f(\vec{r})}\Psi(\vec{r})$.}

To see how a position-dependent phase arises, one may consider a
situation similar to that discussed in Sec. \ref{sec:rearrange}.
That is, consider three-body system consisting of a continuum pair
described by the wavefunction,
$\Psi_{\vec{p}}(\vec{r})=e^{-i\hbar\vec{p}\cdot\vec{r}}C_{\eta_p}(\zeta_p)$
together with a third, mutually interacting particle. Then, upon
breakup, the pair will acquire a local momentum and according to
(\ref{eq:threebody}), the wavefunction will transform as follows:
\begin{equation}
\Psi_{\vec{p}}(\vec{r})\longrightarrow\Psi_{\vec{P}(\vec{r})}(\vec{r})=
e^{-i[\vec{P}(\vec{r})\cdot\vec{r}]}C_{\eta_p}(\zeta_p)=
e^{-i\vec{p}\cdot\vec{r}}e^{+i\vec{D}_{\vec{p}}(\vec{r})\cdot\vec{r}}C_{\eta_p}(\zeta_p).
\end{equation}
Using (\ref{eq:distderived}), one then finds that\footnote{Note
that the ``extra'' $a_0$ arises here because of the scale used in
the plots. i.e., one writes $\vec{r}\rightarrow a_0 r \hat{r}.$},
(for $r\not=0$)
\begin{equation}
\vec{D}_{\vec{p}}(\vec{r})\cdot\vec{r}=a_0\eta_p
\left[1-\cos(\zeta_{p}+2\delta_{\eta_{p}}){}\right]+
ia_0\eta_p\sin(\zeta_{p}+2\delta_{\eta_{p}}).
\end{equation}
Therefore the transformed continuum two-body wavefunction takes the
following form:
\begin{equation}\label{eq:transformpsi}
\Psi_{\vec{P}(\vec{r})}(\vec{r})=e^{i\vec{p}\cdot\vec{r}}
e^{ia_0\eta_p \left[1-\cos(\zeta_{p}+2\delta_{\eta_{p}})\right]}
e^{-a_0\eta_p\sin(\zeta_{p}+2\delta_{\eta_{p}})}C_{\eta_p}(\zeta_p),
\end{equation}
which shows the explicit form for the position dependent phase.
i.e., one may write,
\begin{equation}
S_{\eta_p}(\zeta_p)\equiv ia_0\eta_p
\left[1-\cos(\zeta_{p}+2\delta_{\eta_{p}})\right]
-a_0\eta_p\sin(\zeta_{p}+2\delta_{\eta_{p}}),
\end{equation}
and see that the transformation is indeed indicative of a global
gauge symmetry. eg.,
\begin{equation}
\Psi_{\vec{P}(\vec{r})}(\vec{r})=e^{S_{\eta_p}(\zeta_p)}\Psi_{\vec{p}}(\vec{r})
\end{equation}

It is of course the absolute square of the wavefunction that is
truly important for predicting whether or not this phase will
significantly effect the experimental findings. To this end, one
may take the absolute square of (\ref{eq:transformpsi}) and show
that the \emph{imaginary}-part of the local distortion
contributes a \emph{real}-valued, position-dependent phase. i.e.,
\begin{equation}
\left|\Psi_{\vec{p}}(\vec{r})\right|^2=e^{-2a_0\eta_p\sin(\zeta_{p}+2\delta_{\eta_{p}})}
\left|C_{\eta_p}(\zeta_p)\right|^2=e^{2\Re S_{\eta_p}(\zeta_p)} \left|C_{\eta_p}(\zeta_p)\right|^2.
\end{equation}
The exceedingly small magnitude of the term $2 a_0
\eta_p\sin(\zeta_{p}+2\delta_{\eta_{p}})$ for the relative energy
and reduced masses of the systems considered in current research
findings, makes it clear that one may write
\begin{equation}
e^{-2a_0\eta_p\sin(\zeta_{p}+2\delta_{\eta_{p}})}\longrightarrow 1.
\end{equation}
While this reinforces the fact that the proposed framework will leave the
asymptotic description of the three-body scattering event unaltered, (as was
established in Appendix \ref{app:kc5c}) it does not address the extent to which the
proposed gauge transformation, (\ref{eq:threebody}) will alter the description in
the reaction zone.

To begin an investigation of  the expected behavior in the reaction zone,
one may construct a semi-classical expectation value for the total energy of the
continuum pair, $C_{\eta_p}(\zeta_p)$. To do this recall that the expectation value of
the semi-classical Hamiltonian for this system (before breakup) would be,
\begin{equation}
\left<H_0(\vec{p},\vec{r})\right>=\frac{\hbar^2
p^2}{2\mu}+\frac{Ze^2}{4\pi\epsilon_0}\frac{1}{r}.
\end{equation}
where $p=\frac{\sqrt{2\mu {\mathcal{E}}}}{\hbar}$ is the momentum of the
continuum pair before breakup. During breakup, the pair will acquire a local momentum,
so that the expectation value of the semi-classical Hamiltonian for the system would
then become,
\begin{equation}
\left<H(\vec{P}(\vec{r}),\vec{r})\right>=\frac{\hbar^2
\left(k^2+|\vec{D}_{\vec{p}}(\vec{r})|^2\right)}{2\mu}+\frac{Ze^2}{4\pi\epsilon_0}\frac{1}{r}.
\end{equation}

\begin{figure}
\begin{center}
\fbox{$\text{Three-Body Gauge Variation with ${\mathcal{E}}$}$}\\
\psfrag{r}{$r\ (a_0)$} \psfrag{E}{$\mathcal{E}$} \psfrag{title}{}
\psfrag{mass}{} \psfrag{theta}{} \psfrag{eV}[bc]{Relative Energy}
{\psfrag{H}{$H$} \psfrag{H0}{$H_0$}
\resizebox{2in}{!}{\includegraphics[bb= 0 350 285 540]{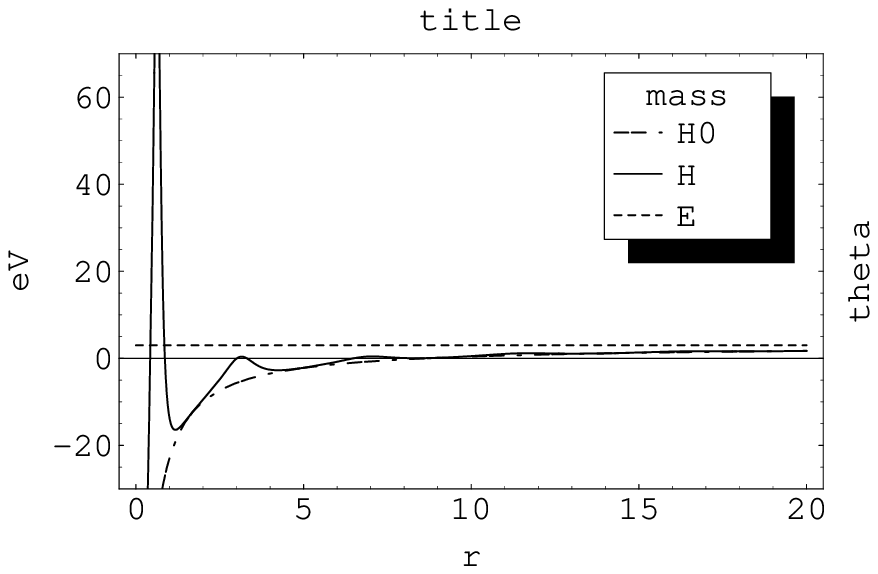}}}\\
(a)\\
{\psfrag{H}{$H_0$} \psfrag{H0}{$H$}
\resizebox{2in}{!}{\includegraphics[bb= 0 350 285 540]{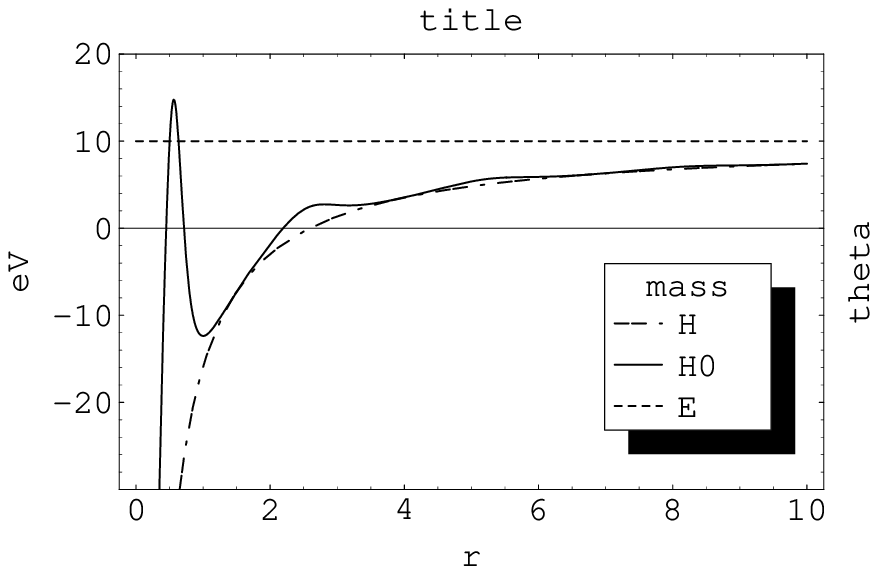}}}\\
(b)\\
{\psfrag{H}{$H_0$} \psfrag{H0}{$H$}
\resizebox{2in}{!}{\includegraphics[bb= 0 350 285 540]{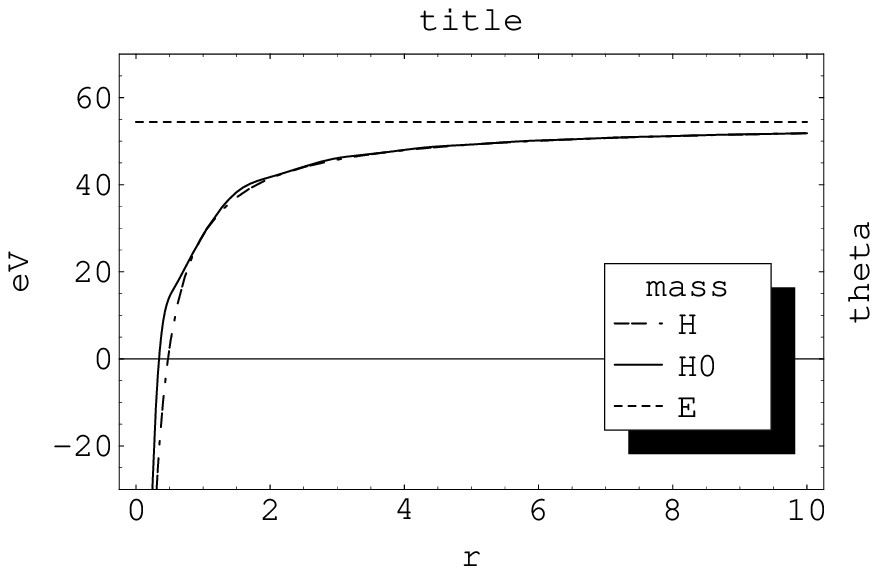}}}\\
(c)
\end{center}
\caption{The effects of the three-body gauge transformation are shown to depend
critically upon the relative energy of the continuum pair. Observe that as the
relative energy of an electron-proton continuum pair, with a scattering angle of
$\theta=0^\circ$, is increased from
(a) ${\mathcal{E}}=3\ \mathrm{eV}$ to
(b) ${\mathcal{E}}=10\ \mathrm{eV}$ to
(c) ${\mathcal{E}}=54.4\ \mathrm{eV}$,
the effects in the reaction zone nearly vanish.}
\label{fig:transformenergy}
\end{figure}

\begin{figure}
\begin{center}
\fbox{$\text{Three-Body Gauge Variation with $\theta$}$}\\
\psfrag{r}{$r\ (a_0)$} \psfrag{H}{$H_0$} \psfrag{H0}{$H$}
\psfrag{E}{$\mathcal{E}$} \psfrag{title}{} \psfrag{mass}{}
\psfrag{theta}{} \psfrag{eV}[bc]{Relative Energy}
{\psfrag{H}{$H_0$} \psfrag{H0}{$H$}
\resizebox{2in}{!}{\includegraphics[bb= 0 350 285 540]{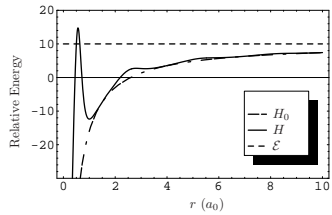}}}\\
(a)\\
{\psfrag{H0}{$H_0$} \psfrag{H}{$H$}
\resizebox{2in}{!}{\includegraphics[bb= 0 350 285 540]{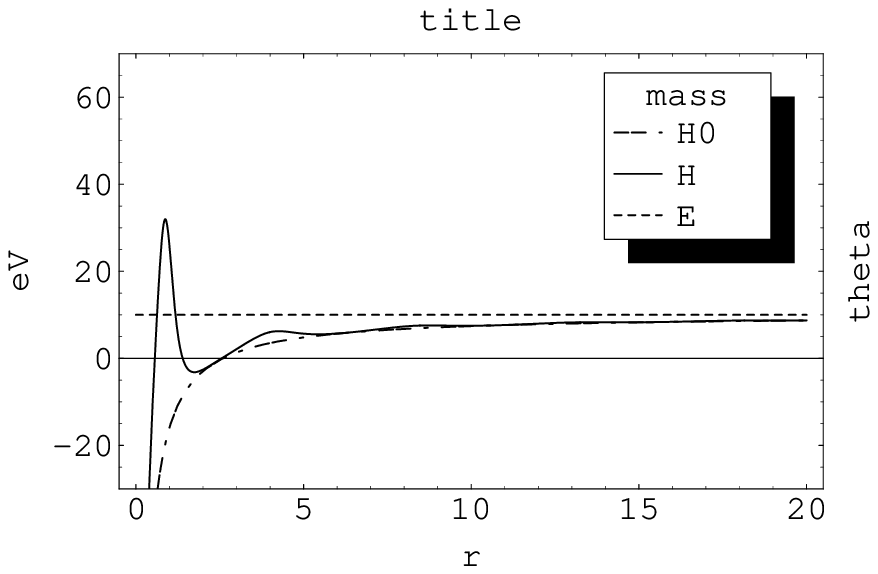}}}\\
(b)\\
{\psfrag{H0}{$H_0$} \psfrag{H}{$H$}
\resizebox{2in}{!}{\includegraphics[bb= 0 350 285 540]{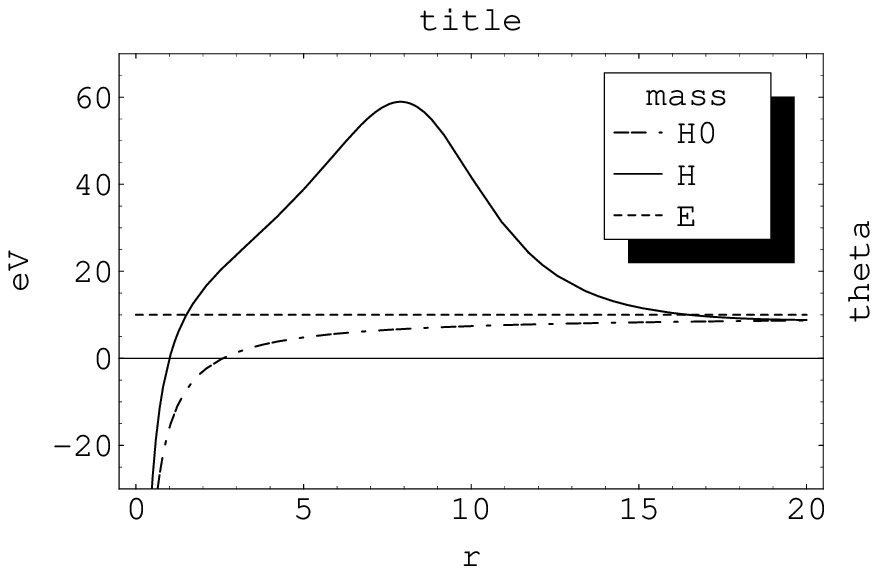}}}\\
(c)
\end{center}
\caption{The effects of the three-body gauge transformation
are shown to depend critically upon the scattering angle of the
continuum pair. Observe that as the scattering angle of an
electron-proton continuum pair, with a relative energy of
${\mathcal{E}}=10\ \mathrm{eV}$, is increased from (a)
$\theta=0^\circ$ to (b) $\theta=75^\circ$ to (c)
$\theta=150^\circ$, the effects in the reaction zone become more
pronounced.} \label{fig:transformangle}
\end{figure}

\begin{figure}
\begin{center}
\fbox{$\text{Three-Body Gauge Variation with $\mu$}$}\\
\psfrag{r}{$r\ (a_0)$} \psfrag{E}{$\mathcal{E}$} \psfrag{title}{}
\psfrag{mass}{} \psfrag{theta}{} \psfrag{eV}[bc]{Relative Energy}
{\psfrag{H}{$H_0$} \psfrag{H0}{$H$}
\resizebox{2in}{!}{\includegraphics[bb= 0 350 285 540]{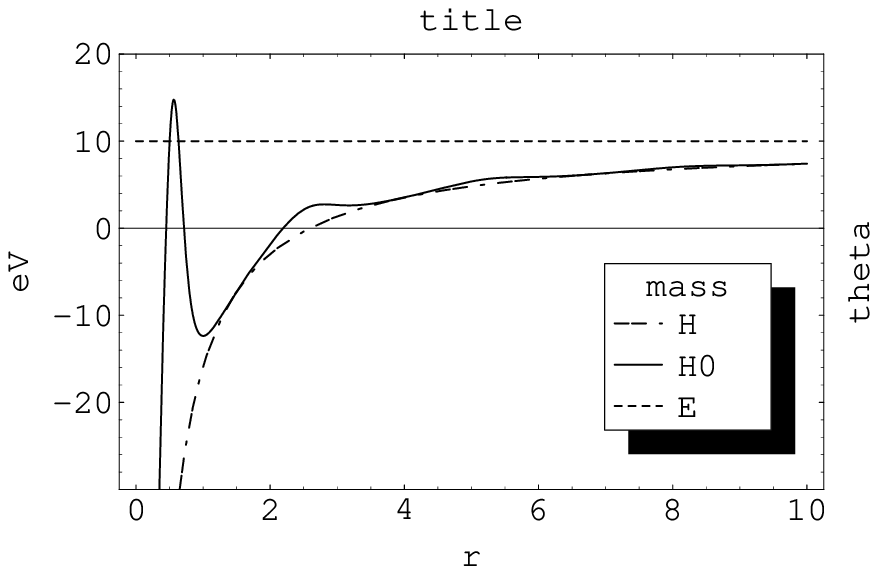}}}\\
(a)\\
{\psfrag{H}{$H$} \psfrag{H0}{$H_0$}
\resizebox{2in}{!}{\includegraphics[bb= 0 350 285 540]{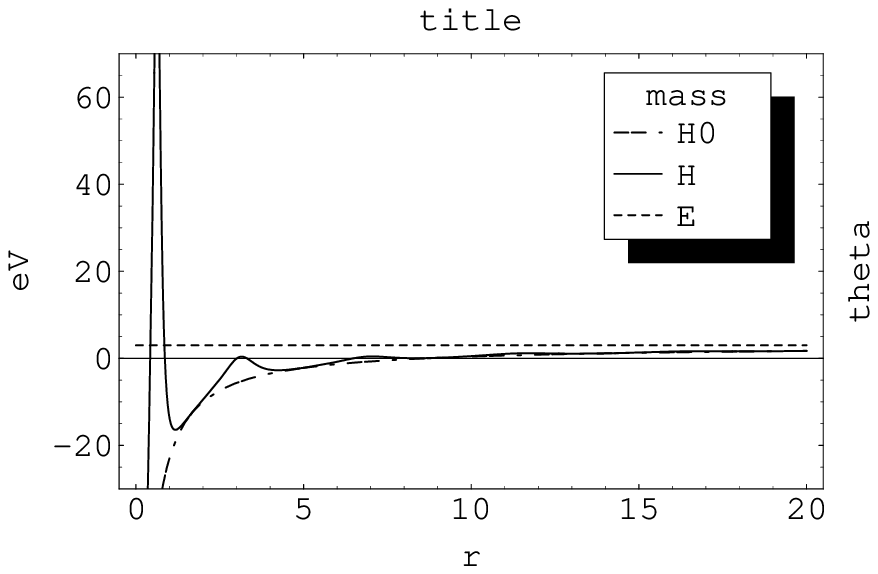}}}\\
(b)\\
{\psfrag{H}{$H_0$} \psfrag{H0}{$H$}
\resizebox{2in}{!}{\includegraphics[bb= 0 350 285 540]{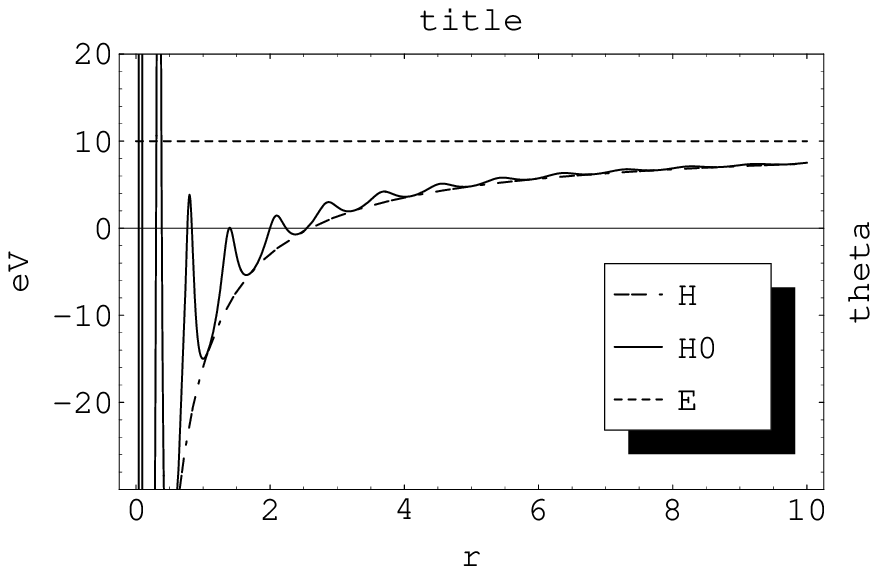}}}\\
(c)
\end{center}
\caption{The effects of the three-body gauge transformation are shown to depend
critically upon the reduced mass of the continuum pair. Observe that as the
reduced mass of the continuum pair, with a relative energy of
${\mathcal{E}}=10\ \mathrm{eV}$ and a scattering angle of $\theta=0^\circ$,
is increased from
(a) $\mu=1\mu_{ep}$ to
(b) $\mu=10\mu_{ep}$ to
(c) $\mu=20\mu_{ep}$,
the effects in the reaction are dramatically altered.}
\label{fig:transformmass}
\end{figure}

The effects of the proposed gauge transformation may then be
observed by plotting the expectation values and varying the
relative energy, ${\mathcal{E}}$ (see Fig.
\ref{fig:transformenergy}), the scattering angle, $\theta$ (see
Fig. \ref{fig:transformangle}), and the reduced mass, $\mu$ (see
Fig. \ref{fig:transformmass}). All of these show undeniably that
while the asymptotic form remains unchanged with respect to
variations of all kinematic parameters, genuine three-body
distortion effects may arise in the reaction zone. Note
specifically that the variation of the relative energy with the
scattering reinforces the interpretation offered in Sec.
\ref{sec:rearrange}. Specifically, the magnitude of the
distortion effects for large scattering angles (corresponding to
the nearly colinear breakup of the continuum pair) in the
reaction zone may cause two opposite-sign-charged particles to
increase their radial separation, and appear to repel one
another. Conversely, two like-sign-charged particles would
decrease their radial separation, and appear to attract one
another.

\begin{figure}
\begin{center}
\fbox{$\text{Asymptotic Phase Variation}$}\\
\psfrag{title}{} \psfrag{phase}{$e^{-2 a_0 \eta_p}$}
{\psfrag{E}{${\mathcal{E}}\ (\mathrm{\mu eV})$} \psfrag{pp}{$p-p$}
\psfrag{pap}{$p-\bar{p}$}
\resizebox{2in}{!}{\includegraphics[bb= 0 300 330 490]{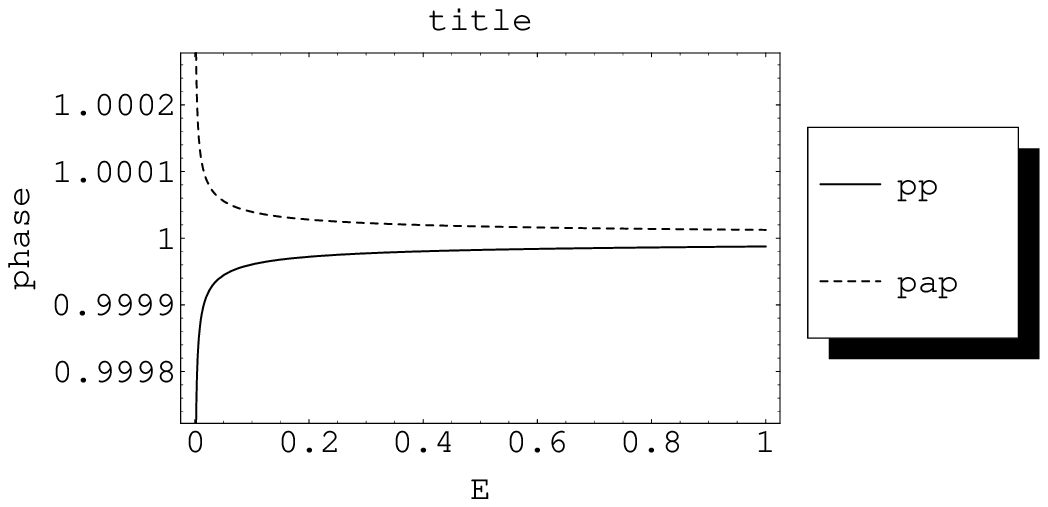}}}\\
(a)\\
{\psfrag{E}{${\mathcal{E}}\ (\mathrm{neV})$} \psfrag{ep}{$e-p$}
\psfrag{aep}{$\bar{e}-p$}
\resizebox{2in}{!}{\includegraphics[bb= 0 300 330 490]{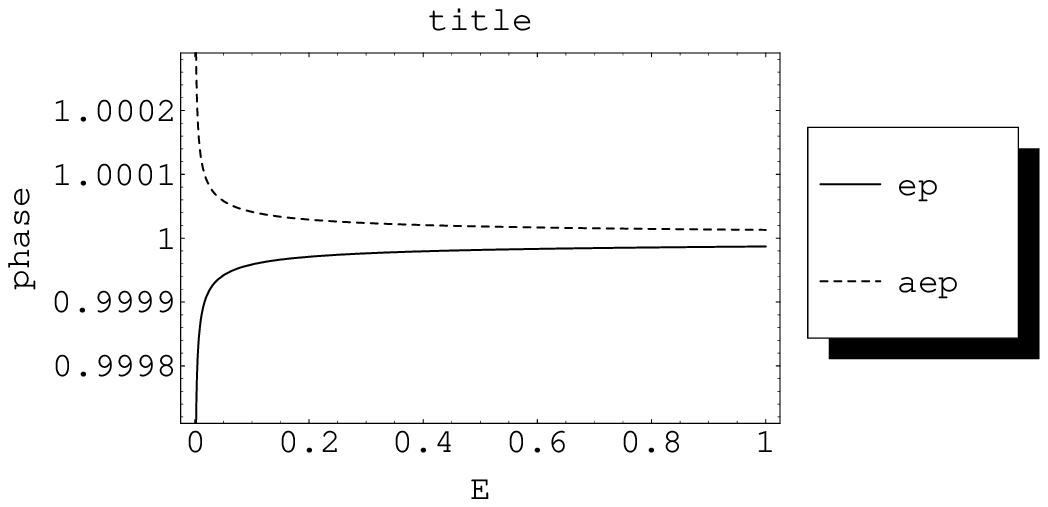}}}\\
(b)\\
{\psfrag{E}{${\mathcal{E}}\ (\mathrm{peV})$} \psfrag{ee}{$e-e$}
\psfrag{aee}{$e-\bar{e}$}
\resizebox{2in}{!}{\includegraphics[bb= 0 300 330 490]{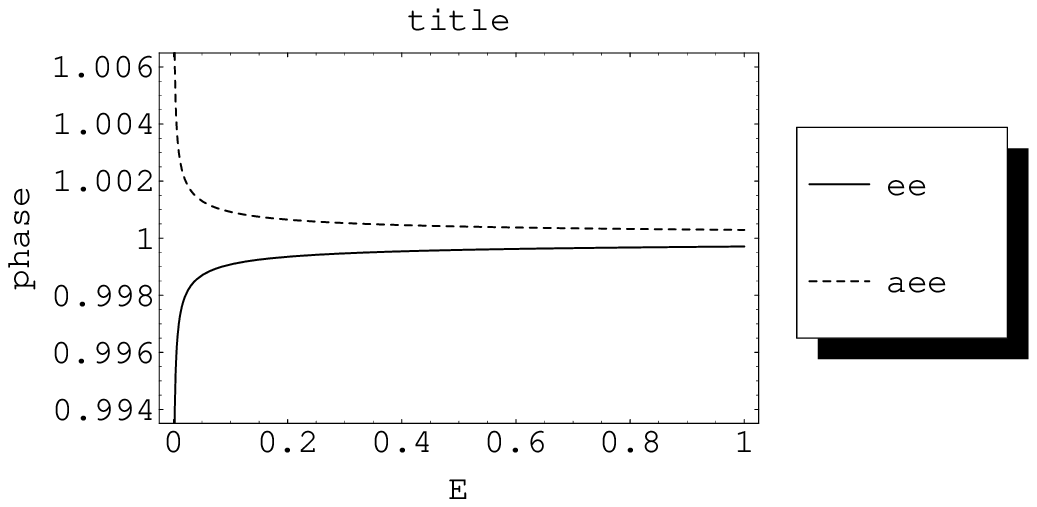}}}\\
(c)
\end{center}
\caption{Shown is the variation of the phase achieved by the
three-body gauge transformation (see text equation
(\ref{eq:threebody})) for (a) a proton-proton or
proton--anti-proton, (b) an electron-proton or positron-proton and
(c) electron-electron or electron-positron continuum pair.
Observe that as the reduced mass of the system is decreased
accordingly, the corresponding energy decrease is sufficient to
detect a variation of one part in a million.}
\label{fig:phasevariation}
\end{figure}

As a last exercise, one may ask if the position-dependent phase introduced by the
three-body gauge formalism could be measured. To answer this question one may observe
that
\begin{equation}
e^{-2\eta_p a_0\sin\left(\zeta_p+2\delta_p\right)}\cong e^{-2\eta_p a_0}=
e^{-\frac{\sqrt{2}z a_0 e^2}{4\pi\epsilon_{0}\hbar}\sqrt{\frac{\mu}{e{\mathcal{E}}}}},
\end{equation}
where (\ref{eq:etap}) has been used for $\eta_p$, and
${\mathcal{E}}$ is measured in electron-volts. As mentioned
above, this contribution approaches unity for all systems of
physical interest in the realm of current experiments in atomic
physics. One can however venture outside of this realm and ask
what energy and/or reduced mass is needed to obtain a measurable
deviation from unity. If one uses the heaviest purely atomic
species, either a proton-proton ($p-p$) or a proton--anti-proton
($p-\bar{p}$) continuum pair, and assumes that a deviation of one
part in a million can be measured, then the energy scale needed
is on the order of $1\ \mu\mathrm{eV}$! The variation of the term
$e^{-2a_0\eta_p}$ at these energies is shown in Figure
\ref{fig:phasevariation}, and illustrates that one may detect a
change in the absolute square of the wavefucntion, thus altering
the relevant scattering amplitude. Moreover, the exceedingly
small energy scale required to detect these asymptotic distortion
effects in either electron(or positron) scattering from hydrogen
or in electron-electron or electron-positron ionization processes
provides a precise understanding for the success achieved in
these areas while ignoring local momentum effects.

\section{Conclusion}
\label{sec:conclusions}

In the above it has been shown that a generalized,
position-dependent local momenta, which depends upon the
congugate coordinate through the logarithmic gradient of a
continuum state Coulomb wave, may provide evidence for the
manifestation of genuine three-body distortion effects in the
reaction zone. The form of this local momenta was derived from a
consideration of the exact three-body wavefunction, for the case
when two of the three, mutually interacting, particles are far
apart, and it indicates that the effects do not depend explicitly
upon the location of the third particle. For this reason, the
effects may be viewed as a distortion of the initial two-body
continuum state wavefunction of the two remaining particles. This
interpretation was adopted, and it was shown that the local
distortion effects could be used to provided a rigorous, physical
description of the observed asymmetry in the data obtained by
Wiese et. al., \cite{Wiese} on the breakup of three massive
Coulomb particles. The degree of this asymmetry was then
predicted with an error of less than $10\%$, and it was also
shown that while the distortion effects were large in the
reaction zone, the asymptotic form of the relevant two-body
interaction may be retained by treating the local momentum
acquisition as a three-body gauge constraint. Furthermore, the
evidence for detecting asymptotic variations in the
$\mu\mathrm{eV}$ range, as presented in Fig.
\ref{fig:phasevariation} suggest that more experimentation be
focused on these low energy, heavy-ion--atom processes. Indeed,
these experiments may yield new insight into a mechanism by which
electrical forces may contribute to the fusion process! That is,
the quantum-mechanical--impulse interpretation offered here shows
that two like-charged particles may in fact attract one another
due to local distortion effects in the reaction zone.

In addition to these findings and predictions, one may learn much
by noting that by adopting the proposed three-body gauge
transformation, one finds that an electron-proton continuum pair
exhibited a very small amount of distortion in the reaction zone.
This result provides a rigorous explanation of why the paradigm 3C
wavefunction works so well for light-ion--atom scattering
\cite{BBK},\cite{Jones}. Moreover, the general framework shows
that an electron-electron continuum pair would experience
distortion effects of lesser magnitude, due to its greatly
decreased reduced mass. These results can again be used to
explain why Qiu et. al.,\cite{Qiu} achieved amazing success in
modeling electron-electron photo-ionization processes, while
ignoring local momentum effects. Indeed, if one recalls the
path-integral interpretation suggested in Sec.
\ref{sec:rearrange}, then the success of these findings for
light-ion--atom processes may be attributed to the fact that the
leading contribution to the relevant cross sections are the
``paths'' along which the local distortion effects are
identically zero. While no analytic form exists for calculation
of these roots, one can compile a table for use in numerical
calculations and construct a solution that better reflects the
physical nature of the three-body interaction.


\appendix

\section{The Logarithmic Derivative of the ${}_{1}F_{1}(\lowercase{i\eta_p,1;-1\zeta_p})$ Function}
\label{app:1f1}

The logarithmic derivative of the confluent hypergeometric
function is defined here as follows:
\begin{eqnarray}
\vec{D}_{\vec{p}}(\vec{r})&\equiv&i\vec{\nabla}_r\ln{_1F_1\left(i\eta_p,1;-i\zeta_p\right)}
=i\frac{\vec{\nabla}_{r}{}_{1}F_{1}(i\eta_p,1;-i\zeta_p)}{{}_{1}F_{1}(i\eta_p,1;-i\zeta_p)}\nonumber\\&
&\\
&=&i\eta_p{}p\frac{{}_{1}F_{1}\left[(i\eta_p+1),2;-i\zeta_p\right]}{{}_{1}F_{1}(i\eta_p,1;-i\zeta_p)}(\hat{p}+\hat{r}).\nonumber
\end{eqnarray}
Then using the well known recursion relationship,(See reference \cite{GR} for
instance.)
\begin{equation}\label{eq:recur}
\frac{z}{b}{}_1F_1\left[(a+1),(b+1),z\right]={}_1F_1\left[(a+1),b,z\right]-{}_1F_1\left(a,b,z\right),
\end{equation}
one may write\footnote{Because of the division by $\zeta_p$ in equation
(\ref{eq:dapp}), the region of validity for the logarithmic derivative is limited to
$\zeta_p>0$ only as indicated. One can however find the actual value at $\zeta_p=0$.
\emph{c.f.} equation (\ref{eq:dac}).}

\begin{eqnarray}\label{eq:dapp}
\vec{D}_{\vec{p}}(\vec{r})&=&i\eta_p{p}\frac{\left[{}_{1}F_{1}\left[(i\eta_p+1),1;-i\zeta_p\right]-{}_{1}F_{1}(i\eta_p,1;-i\zeta_p)\right]}{\zeta_p
{}_{1}F_{1}(i\eta_p,1;-i\zeta_p)} (\hat{p}+\hat{r}),\quad\quad\quad\zeta_p>0.
\end{eqnarray}

Further simplification of the logarithmic derivative is obtained
by use of the Kummer relation\cite{GR} for the confluent
hypergeometric equation. \emph{e.g.},
\begin{equation}
{}_1F_1\left(a,b,z\right)=e^z{}_1F_1\left[(b-a),b,-z\right].
\end{equation}
With this relation one finds that the distortion may be rewritten as follows:
\begin{subequations}
\begin{eqnarray}
\vec{D}_{\vec{p}}(\vec{r})&=&i\eta_p{p}\frac{\left[e^{-i\zeta_p}{}_{1}F_{1}(-i\eta_p,1;i\zeta_p)-{}_{1}F_{1}(i\eta_p,1;-i\zeta_p)\right]}{\zeta_p\
{}_{1}F_{1}(i\eta_p,1;-i\zeta_p)} (\hat{p}+\hat{r})\\
&=&\frac{\eta_p}{r\
\hat{p}\cdot(\hat{p}+\hat{r})}
\left[1-e^{-i\zeta_p}\frac{{}_{1}F_{1}(-i\eta_p,1;i\zeta_p)}{{}_{1}F_{1}(i\eta_p,1;-i\zeta_p)}\right](\hat{p}+\hat{r}),\quad\quad\quad
{}r>0.
\label{eq:ratio}
\end{eqnarray}
\end{subequations}

At this point note that the ratio of confluent hypergeometric
equations in equation (\ref{eq:ratio}) is in fact a very special
case because the function on top is the complex conjugate of the
function on the bottom! i.e.,
\begin{equation}
{}_{1}F_{1}(-i\eta_p,1;i\zeta_p)={{}_{1}F_{1}}^{\dag}(i\eta_p,1;-i\zeta_p),
\end{equation}
so that one may define a \emph{real}-valued, position-dependent phase,
\begin{equation}
\tan\left[\delta_{\eta_p}(\zeta_p)\right]=\frac{\Im\left[{}_{1}F_{1}(i\eta_p,1;-i\zeta_p)\right]}
{\Re\left[{}_{1}F_{1}(i\eta_p,1;-i\zeta_p)\right]}.
\end{equation}

The logarithmic derivative may then be written as follows:
\begin{equation}\label{eq:dac}
\vec{D}_{\vec{p}}(\vec{r})=
\Bigg\{
\begin{tabular}{c}
$i\eta_p{p}(\hat{p}+\hat{r}), \quad r=0$\\
$\frac{\eta_p}{r}\left[1-e^{-i(\zeta_p+2\delta_{\eta_p}(\zeta_p))}\right]
\frac{(\hat{p}+\hat{r})}{\hat{p}\cdot(\hat{p}+\hat{r})}, \quad r>0$
\end{tabular},
\end{equation}
and one sees that in this form, the local distortion is analytic everywhere,
except possibly at $r=0$. This result is very significant, because it allows one to
investigate the nature of the local momenta well inside of the interior regions of
a scattering event in a rigorous manner. Only in this way can one determine the
relative importance of these local effects.

\section{An Alternative Three-Body Wavefunction}
\label{app:kc5c}

To solve equation (\ref{eq:chi}) for the unknown function
$\chi_{\vec{k}_{\alpha},\vec{q}_{\alpha}}(\vec{r}_{\alpha},\vec{\rho}_{\alpha})$
one normally assumes that the interaction energy (cf, equation (\ref{eq:myenergy})) satisfies
$${\mathcal{E}_I}=0.$$
This technique was developed by Popalilios as referenced in
\cite{BBK}, and asserts that the total center of mass energy of the three-body system
is partitioned among the two reduced mass clusters, $\mu_\alpha$ and $M_\alpha$.
Hence one may write
\begin{equation}
\left[-\frac{\hbar^2}{2\mu_\alpha}\vec{\nabla}^2_{\vec{r}_\alpha}-
i\frac{\hbar}{\mu_\alpha}\vec{K}(\vec{r}_\alpha)\cdot\vec{\nabla}_{\vec{r}_\alpha}-
\frac{\hbar^2}{2M_\alpha}\vec{\nabla}^2_{\vec{\rho}_\alpha}-
i\frac{\hbar}{M_\alpha}\vec{Q}(\vec{\rho}_\alpha)
\cdot\vec{\nabla}_{\vec{\rho}_\alpha}+
V_\alpha^I(\vec{r}_\alpha,\vec{\rho}_\alpha)\right]
\chi_{\vec{k}_{\alpha},\vec{q}_{\alpha}}(\vec{r}_{\alpha},\vec{\rho}_{\alpha})=0.
\end{equation}
and see that $\chi_{\vec{k}_{\alpha},\vec{q}_{\alpha}}(\vec{r}_{\alpha},\vec{\rho}_{\alpha})$
must be built to incorporate each term in the interaction potential.

To accomplish this, assume that $\chi_{\vec{k}_{\alpha},\vec{q}_{\alpha}}(\vec{r}_{\alpha},\vec{\rho}_{\alpha})$
is given by
\begin{equation}
\chi_{\vec{k}_{\alpha},\vec{q}_{\alpha}}(\vec{r}_{\alpha},\vec{\rho}_{\alpha})=
C_{-\eta_{q_\alpha}}(\zeta_{q_\alpha})f_\beta(\vec{r}_{\alpha},\vec{\rho}_{\alpha})
f_\gamma(\vec{r}_{\alpha},\vec{\rho}_{\alpha}),
\end{equation}
where $C_{-\eta_{q_\alpha}}(\zeta_{q_\alpha})$ satisfies
\begin{equation}
\left[-\frac{\hbar^2}{2M_\alpha}\vec{\nabla}^2_{\vec{\rho}_\alpha}-
i\frac{\hbar}{M_\alpha}\vec{q}_\alpha\cdot\vec{\nabla}_{\vec{\rho}_\alpha}
-\frac{Z_\alpha(Z_\beta+Z_\gamma) e^2}{4 \pi
\epsilon_0}\frac{1}{\rho_\alpha}\right]
C_{\eta_{q_\alpha}}(\zeta_{q_\alpha})=0.
\end{equation}

Because the function $C_{-\eta_{q_\alpha}}(\zeta_{q_\alpha})$
is a known function, which exactly incorporates the interaction
term $-\frac{Z_\alpha(Z_\beta+Z_\gamma)e^2}{4\pi \epsilon_0
\rho_\alpha}$, this substitution results in the following (exact) coupled
equations for the functions $f_\nu(\vec{r}_\nu,\vec{\rho}_\nu)$:($\nu=\beta, \gamma$)
\begin{subequations}\label{eq:myexact}
\begin{eqnarray}
& \resizebox{5in}{!}{$
\left[-\frac{\hbar^2}{2\mu_\alpha}\vec{\nabla}^2_{\vec{r}_\alpha}- \frac{\hbar^2}{2
M_\alpha}\vec{\nabla}^2_{\vec{\rho}_\alpha}-
i\frac{\hbar}{\mu_\alpha}\vec{K}_\alpha^\gamma(\vec{r}_{\alpha};\vec{r}_\gamma,\vec{\rho}_\gamma)\cdot\vec{\nabla}_{\vec{r}_\alpha}
-i\frac{\hbar}{M_\alpha}\vec{Q}_\alpha^\gamma(\vec{\rho}_\alpha,\vec{r}_\gamma,\vec{\rho}_\gamma)\cdot\vec{\nabla}_{\vec{\rho}_\alpha}+
V_{\nu,\mathrm{eff}}^{(\gamma)}(\vec{r}_\beta,\vec{\rho}_\alpha)
\right]f_\beta(\vec{r}_\beta,\vec{\rho}_\beta)=0 $}
,&\\
& \resizebox{5in}{!}{$
\left[-\frac{\hbar^2}{2 \mu_\alpha}\vec{\nabla}^2_{\vec{r}_\alpha}-
\frac{\hbar^2}{2 M_\alpha}\vec{\nabla}^2_{\vec{\rho}_\alpha}-
i\frac{\hbar}{\mu_\alpha}\vec{K}_\alpha^\beta(\vec{r}_{\alpha};\vec{r}_\beta,\vec{\rho}_\beta)\cdot\vec{\nabla}_{\vec{r}_\alpha}
-i\frac{\hbar}{M_\alpha}\vec{Q}_\alpha^\beta(\vec{\rho}_\alpha,\vec{r}_\beta,\vec{\rho}_\beta)\cdot\vec{\nabla}_{\vec{\rho}_\alpha}+
V_{\gamma,\mathrm{eff}}^{(\beta)}(\vec{r}_\gamma,\vec{\rho}_\alpha)
\right]f_\gamma(\vec{r}_\gamma,\vec{\rho}_\gamma)=0 $},&
\end{eqnarray}
\end{subequations}
where (for $\epsilon\not=\nu$, with $\epsilon=\beta,\gamma$)
\begin{subequations}
\begin{eqnarray}
\vec{K}_\alpha^\epsilon(\vec{r}_{\alpha};\vec{r}_\epsilon,\vec{\rho}_\epsilon)&=&
\vec{k}_\alpha-i\vec{\nabla}_{\vec{r}_\alpha}\ln C_{\eta_{k_\alpha}}(\zeta_{k_\alpha})
-\frac{i}{2}\vec{\nabla}_{\vec{r}_\alpha}\ln f_\epsilon(\vec{r}_\epsilon,\vec{\rho}_\epsilon), \\
\vec{Q}_\alpha^\epsilon(\vec{\rho}_{\alpha};\vec{r}_\epsilon,\vec{\rho}_\epsilon)&=&
\vec{q}_\alpha-i\vec{\nabla}_{\vec{\rho}_\alpha}
\left[\ln
C_{\eta_{q_\alpha}}(\zeta_{q_\alpha})C_{-\eta_{q_\alpha}}(\zeta_{q_\alpha})\right]
-\frac{i}{2}\vec{\nabla}_{\vec{\rho}_\alpha}\ln f_\epsilon(\vec{r}_\epsilon,\vec{\rho}_\epsilon),
\end{eqnarray}
\end{subequations}
and\footnote{Note that the general form of the local distortion given in (\ref{eq:distderived}) has
been used to derive this form.}
\begin{equation}
V_{\nu,\mathrm{eff}}^{(\epsilon)}(\vec{r}_\nu,\vec{\rho}_\alpha)=
\frac{Z_\alpha Z_\epsilon e^2}{4 \pi \epsilon_0}\frac{1}{r_\nu}\times
\left\{1+\frac{4 \pi \epsilon_0 \eta_{q_\alpha}^2}{Z_\alpha Z_\epsilon e^2 M_\alpha}
\left(\frac{r_\nu}{\rho_\alpha}\right)
\left[1-e^{-i(\zeta_{q_\alpha}+2\delta_{\eta_{q_\alpha}}(\zeta_{q_\alpha}))}\right]
\left[1-e^{-i(\zeta_{q_\alpha}+2\delta_{-\eta_{q_\alpha}}(\zeta_{q_\alpha}))}\right]\right\}.
\end{equation}

While these equations may seem even more complicated than the
original equation, there are many important aspects to note:
\begin{itemize}
\item The effective potential reduces to the Coulomb potential for the $\nu$-channel,
in the asymptotic regions $\Omega_0$ and $\Omega_\alpha$. i.e.,
\begin{equation}
V_{\nu,\mathrm{eff}}^{(\epsilon)(\Omega_0,\Omega_\alpha)}(\vec{r}_\nu,\vec{\rho}_\alpha)=
\frac{Z_\alpha Z_\epsilon e^2}{4 \pi \epsilon_0}\frac{1}{r_\nu}
\end{equation}
\item The second term in
$\vec{Q}_\alpha^\epsilon(\vec{\rho}_\alpha,\vec{r}_\epsilon,\vec{\rho}_\epsilon)$
vanishes in the asymptotic regions because,
\begin{equation}\label{eq:cancel}
\left[C_{\eta_{q_\alpha}}(\zeta_{q_\alpha})C_{-\eta_{q_\alpha}}(\zeta_{q_\alpha})
\right]_{\lim_{{\rho_\alpha}\to\infty}}\longrightarrow
e^{-i\eta_{q_\alpha}\ln \zeta_{q_\alpha}}e^{+i\eta_{q_\alpha}\ln \zeta_{q_\alpha}}=1.
\end{equation}
(Indeed, this cancellation of the logarithmic phases was ``built in'' to
the solution!)
\item The first two terms in $\vec{K}_\alpha^\epsilon(\vec{r}_\alpha,\vec{r}_\epsilon,\vec{\rho}_\epsilon)$
may be identified as a position-dependent, local
momentum. i.e.,
$$\vec{K}_\alpha(\vec{r}_\alpha)=\vec{k}_\alpha-
i\vec{\nabla}_{\vec{r}_\alpha} \ln C_{\eta_{k_\alpha}}(\zeta_{k_\alpha}).$$
\item The solutions are coupled in a completely symmetric way by the
terms $-\frac{i}{2}\vec{\nabla}_{\vec{r}_\alpha}\ln
f_\nu(\vec{r}_\epsilon,\vec{\rho}_\epsilon)$ and \\
$-\frac{i}{2}\vec{\nabla}_{\vec{\rho}_\alpha}\ln
f_\nu(\vec{r}_\epsilon,\vec{\rho}_\epsilon)$, so that a numerical
solution of these equations would be manifestly less
computationally intensive.
\end{itemize}
Furthermore, because the equations (\ref{eq:myexact}) are exact, they would be more reliable
in \emph{ab initio} calculations.

Approximate solutions that are valid through second order, may
be achieved in a direct manner following the techniques outlined in \cite{Engelns} and may be
written in the form of distorted and coupled Coulomb waves. e.g.,
\begin{equation}
f_{\nu}(\vec{r}_\nu,\vec{\rho}_\nu)=C_{\eta_{K_\nu^\epsilon}}(\zeta_{K_\nu^\epsilon})\equiv
\Gamma(1-i\eta_{K_\nu^\epsilon})e^{-\frac{\pi
\eta_{K_\nu^\epsilon}}{2}}{}
_1F_{1}\left(i\eta_{K_\nu^\epsilon},1,-i\zeta_{K_\nu^\epsilon}\right),
\end{equation}
where the conventional, two-body normalization procedure has been employed so that
that incoming wave has unit magnitude.

The complete three-body wavefunction may then be written in the
form:
\begin{equation}
\Psi_{\vec{k}_\alpha,\vec{q}_\alpha}^{KC5C}
(\vec{r}_\alpha,\vec{\rho}_\alpha)=
e^{i(\vec{k}_\alpha\cdot\vec{r}_\alpha+\vec{q}_\alpha\cdot\vec{\rho}_\alpha)}
C_{\eta_{k_\alpha}}(\zeta_{k_\alpha})
C_{\eta_{q_\alpha}}(\zeta_{q_\alpha})C_{-\eta_{q_\alpha}}(\zeta_{q_\alpha})
C_{\eta_{K_\beta^\gamma}}(\zeta_{K_\beta^\gamma})
C_{\eta_{K_\gamma^\beta}}(\zeta_{K_\gamma^\beta})
\end{equation}
a product of five, kinematically coupled Coulomb waves. To see
that this solution is valid, first note that one can show in a
straightforward manner that the distorted Coulomb waves have the following
asymptotic form:
\begin{equation}
C_{\eta_{K_\nu^\epsilon}}^{(\Omega_0)}(\zeta_{K_\nu^\epsilon})
\rightarrow C_{\eta_{k_\nu}}(\zeta_{k_\nu})
\end{equation}
Hence, due to the relationship (\ref{eq:cancel}), the asymptotic form in $\Omega_0$ will be
identical to that of the Redmond, 3C wavefunction.

\newpage

\begin{thebibliography}{}
\bibitem{AM}E. O. Alt and A. M. Mukhamedzhanov, Phys. Rev. A $\mathbf{47}$, 2004(1993).
\bibitem{Wiese} L. Wiese \emph{et. al.}, PRL $\mathbf{79}$, 4982
(1997).
\bibitem{Fiziev} P.P Fiziev, T. Y. Fizieva, Few-Body Systems $\mathbf{2}$ 71 (1987).
\bibitem{Redmond} P. J. Redmond, ca. 1972,unpublished (as referenced by Rosenberg).
\bibitem{Rosenberg} L. Rosenberg, Phys. Rev. D $\mathbf{8}$, 1833(1972).
$\mathbf{56}$, 370(1997).
\bibitem{BBK}M. Brauner \emph{et. al.}, J. Phys. B $\mathbf{22}$, 2265(1989).
\bibitem{GR} I. S. Gradshteyn and I. M. Ryzhik, \underline{Table of Integrals, Series, and Products}.
Academic Press, London (1965).
\bibitem{Berakdar} J. Berakdar, J. S. Briggs, Phys. Rev. A
\bibitem{Jones} S. Jones and D. Madison, Phys. Rev. A
$\mathbf{55}$ 444 (1997).
\bibitem{Qiu} Y. Qiu \emph{et. al.}, Phys. Rev. A $\mathbf{57}$,
R1489 (1998).
\bibitem{Lieber} M. Lieber and A. M. Mukhamedzhanov, Phys. Rev. A $\mathbf{54}$, 3078(1996).
\bibitem{Engelns}A. Engelns \emph{et. al,}, J. Phys. B $\mathbf(30)$, L811(1997).

\end{thebibliography}
\end{document}